\documentclass[11pt,reqno]{amsproc}

\usepackage{fullpage}
\usepackage{graphicx}
\usepackage{subfig}
\usepackage{wrapfig}
\usepackage{cite}
\usepackage{tabu}
\usepackage{multirow}
\usepackage{color}
\usepackage{hyperref}
\graphicspath{ {./Figures/}}
\begin{document}
\title{Comparative study of finite element methods using the Time-Accuracy-Size (TAS) spectrum analysis}

\author{Justin Chang}
\address{Computational and Applied Mathematics, Rice University}
%\curraddr{}
\email{jychang48@rice.edu}
%\thanks{}

\author{Maurice S. Fabien}
\address{Computational and Applied Mathematics, Rice University}
%\curraddr{}
\email{fabien@rice.edu}
%\thanks{}

\author{Matthew G. Knepley}
\address{Computer Science and Engineering, University at Buffalo}
%\curraddr{}
\email{knepley@buffalo.edu}
%\thanks{}

\author{Richard T. Mills}
\address{Computer Science and Mathematics Division, Argonne National Laboratory}
%\curraddr{}
\email{rtmills@anl.gov}
%\thanks{}

\begin{abstract}
We present a performance analysis appropriate for comparing algorithms 
using different numerical discretizations. By taking into account the 
total time-to-solution, numerical accuracy with respect to an error 
norm, and the computation rate, a cost-benefit analysis can be performed 
to determine which algorithm and discretization are particularly suited
for an application. This work extends the performance spectrum model 
in~\cite{ChangNakshatralaKnepleyJohnsson2017} for interpretation of hardware and 
algorithmic tradeoffs in numerical PDE simulation. As a proof-of-concept, 
popular finite element software packages are used to illustrate this 
analysis for Poisson's equation.

\end{abstract}

\maketitle

\section{Introduction}
Computational scientists help bridge the gap between theory 
and application by translating mathematical techniques into robust software. One of
the most popular approaches undertaken is the development of sophisticated 
finite element packages like FEniCS/Dolfin~\cite{Logg2009,alnaes2015fenics}, deal.II~\cite{bangerth2016deal,AgelekAndersonBangerthBarth2017}, 
Firedrake~\cite{rathgeber2016firedrake}, LibMesh~\cite{libMeshPaper}, and 
MOOSE~\cite{Gaston2009} which provide application scientists the necessary
scientific tools to quickly address their specific needs. Alternatively, stand alone finite element
computational frameworks built on top of parallel linear algebra libraries like 
PETSc~\cite{petsc-user-ref,petsc-web-page} may need to be developed to 
address specific technical problems such as enforcing maximum principles 
in subsurface flow and transport modeling~\cite{Chang_JOMP_2017,
Chang_CMAME_2017,Mapakshi_JCP_2018} or modeling atmospheric and other geophysical
phenomena~\cite{brown2013icesheet,may_ptatin}, all of which could require field-scale 
or even global-scale resolutions. As scientific problems grow increasingly 
complex, the software and algorithms used must be fast, scalable, and 
efficient across a wide range of hardware architectures and scientific applications,
and new algorithms and numerical discretizations may need to be introduced. 
The ever increasing capacity and sophistication of processors, memory systems, 
and interconnects bring into question not only the performance of these new 
techniques, but their feasibility for large-scale problems. Specifically, how scalable is the software in both 
the algorithmic and parallel sense? Difficult problems require highly accurate numerical 
solutions, so it is desirable to take into consideration the solution accuracy 
along with both hardware utilization and algorithmic scalability. This paper is concerned with 
benchmarking the performance of various scientific software using analytic techniques.

\subsection{Overview of scaling analyses}
% Strong
% How does my solver handle one problem?
The most basic parallel scaling analysis, known as \textit{strong-scaling}, looks at the 
marginal efficiency of each additional processor working on a given problem~
\cite{Amdahl1967,Eijkhout2014,Knepley2017}. A series of experiments is
run using a fixed problem size but varying the number of processors used. It is typical to 
plot the number of processors $P$ against the speedup, defined as the time on one 
processor divided by the time on $P$ processors. Perfect speedup would result in a curve of 
unit positive slope. Because application scientists often have reason to solve a problem at a 
given resolution and spatial extent, strong-scaling is often of most interest: They may want to 
solve their problems in as little wall-clock time as possible, or, when running on 
shared resources, to complete a set of simulations in reasonable time without using 
too many CPU hours of their allocation (running in a strong-scaling ``sweet spot'' for 
their problem and machine). A simple strong-scaling analysis may, however, make it 
difficult to disentangle sources of inefficiency (serial sections,
communication, latency, algorithmic problems, etc.). 

% Weak
% How does my solver handle a range of problem sizes?
In some cases, application scientists may be interested in exploring a problem 
at a range of resolutions or spatial extents. This leads naturally to {\it weak-scaling}~
\cite{Gustafson1988,Eijkhout2014,Knepley2017} scenarios: 
Instead of fixing the global problem size, they fix the portion of the problem on each 
processor and scale the entire problem size linearly with the number of processors. 
It is typical to plot the number of processors $P$ against the efficiency, defined as 
the time on one processor over the time on $P$ processors. Perfect efficiency would 
result in a flat curve at unity. This analysis shows the marginal efficiency of adding another 
subproblem, rather than just a processor, and separates communication overhead and
algorithmic inefficiency from the problem of serial sections.

% Static
% How does my machine handle a range of problem sizes?
It is difficult to see, from either strong- or weak-scaling diagrams, how a given machine or 
algorithm will handle a variety of workloads. For example, is there is a minimum solution 
time where solver operations are swamped by latency? Is there a problem size where 
algorithmic pieces with suboptimal scaling start to dominate? We can examine these 
questions by running a series of problem sizes at fixed parallelism, called 
\textit{static-scaling}~\cite{Brown2016,ChangNakshatralaKnepleyJohnsson2017}. It is typical to 
plot the computation rate, in number of degrees-of-freedom (DoF) per time, against the time. 
Perfect scaling would result in a flat curve. Tailing off at small times is generally due to 
latency effects, and the curve will terminate at the smallest turnaround time for the 
machine. Decay at large times indicates suboptimal algorithmic
performance or suboptimal memory access patterns and cache misses for 
larger problems. Thus we can see both strong- and weak-scaling
effects on the same graph. It is also harder to game the result, since 
runtime is reported directly and extra work is directly visible. Static-scaling is a useful 
analytic technique for understanding the performance and scalability of
PDE solvers across different hardware architectures and software implementations.

From the standpoint of scientific computing, a significant drawback of all of the above 
types of analyses is that they treat all computation equally and do not consider the theoretical
convergence rate of the particular numerical discretization. The floating-point operations 
(FLOPs) done in the service of a quadratically convergent method, for instance, should 
be considered more valuable than those done for a linear method if both are in the basin of
convergence. These analyses do not depend at all on numerical accuracy so the actual 
convergence behavior is typically measured empirically using the Method of Manufactured 
Solutions (MMS).  If all equations are created equal, say for methods with roughly similar 
convergence behavior, then static-scaling is viable. However, it is insufficient 
when comparing methods with very different approximation properties. Any comparative 
study between different finite element methods or numerical
discretization should factor accuracy into the scaling analyses.

\subsection{Main contribution}
The aim of this paper is to present an alternative performance spectrum model which takes 
into account time-to-solution, accuracy of the numerical solution, and size of the problem hence
the Time-Accuracy-Size (TAS) spectrum analysis. These are the three metrics of most importance
when a comparative study involving different finite element or any numerical methods is needed.
Not every DoF has an equal contribution to the discretization's level of accuracy, so the DoF per 
time metric alone would be neither a fair nor accurate way of assessing the quality of a particular 
software's implementation of the finite element method. An outline of the salient features of this paper
are listed below:
\begin{itemize}
\item We provide a modification to the static-scaling analysis incorporating numerical accuracy.
\item We present the TAS spectrum and discuss how to analyze its diagrams.
\item Popular software packages, such as FEniCS/Dolfin, deal.II, Firedrake, and PETSc, are compared using the Poisson problem.
\item Different single-field finite element discretizations, such as the Galerkin and Discontinuous Galerkin methods, are also compared.
\item The analysis is extended to larger-scale computations, i.e.\ over 1K MPI processes.
\end{itemize}

The rest of the paper is organized as follows. In Section~\ref{sec:spectrum}, we 
present the framework of the Time-Accuracy-Size (TAS) spectrum model and outline 
how to interpret the diagrams. In Section~\ref{sec:theory}, we provide a theoretical 
derivation of the TAS spectrum and provide example convergence plots one may expect
to see. In Section~\ref{sec:setup}, we describe in detail how the finite element
experiments are setup. In Section~\ref{sec:results}, we demonstrate various
ways the TAS spectrum is useful by running various test cases. Conclusions 
and possible extensions of this work are outlined in 
Section~\ref{sec:conclusion}.

\section{TAS Spectrum}\label{sec:spectrum}
In order to incorporate accuracy into our performance analysis, 
we must first have an idea of convergence, or alternatively the numerical 
accuracy of a solution. In this paper, we will measure
solution accuracy using the $L_2$ norm of the error $err$,
\begin{align}
  err  = \left\lVert u_h - u \right\rVert_{L_2},
\end{align}
where $u_h$ is the finite element solution, $u$ is the exact solution,
and $h$ is some measure of our resolution such as the longest edge in any mesh element. 
Convergence means that our error shrinks as we increase our resolution, 
so that $\lim_{h\to0} err = 0$. In fact, we expect most discretization methods 
to have a relation of the form
\begin{align}
 err \le C h^\alpha,
\end{align}
where $C$ is some constant and $\alpha$ is called the \textit{convergence rate} of the method. 
This relation can be verified by plotting the logarithm of the resolution $h$ against the negative 
logarithm of the error $err$, which we call the \textit{digits of accuracy}
(DoA),
\begin{align}
  \mathrm{DoA} = -\log_{10} err.
\end{align}
This should produce a line with slope $-\alpha$, and is typically called a \textit{mesh convergence}
diagram~\cite{BrennerScott02,Knepley2017}. If we use the number of unknowns $N$ (or DoF) instead of the 
resolution $h$, the slope of the line will be modified. For most schemes $N = D h^{-d}$, where $d$ 
is the spatial dimension and $D$ is a constant, so that the slope would become $\alpha/d$. In 
our development, we will use this form of convergence diagram and call $\log_{10} N$ the 
\textit{digits of size} (DoS). Much as weak-scaling explains the behavior of algorithm on a 
range of problem sizes, the mesh-convergence diagram explains the behavior of a discretization.

% Efficacy
In order to incorporate accuracy information, we will imitate the static-scaling analysis by examining the rate of
accuracy production. We will introduce a measure called efficacy, defined to be error multiplied by time. Smaller efficacy
% RTM: I changed things so that we call efficacy a "measure", rather than a "unit". I think this is the correct terminology.
% For instance, I'd say that "velocity" is a measure -- defined as distance over time -- but there are different units of 
% measure for reporting velocity: meters per second, miles per hour, furlongs per fortnight, etc.
is desirable, as this means either smaller error or smaller time. We introduce the \textit{digits of efficacy} (DoE) as
the logarithm of error multiplied by time. As shown in Section~\ref{sec:theory}, this rate has a linear dependence on
problem size, and slope $d - \alpha$. Our \textit{accuracy scaling} analysis plots digits of efficacy against time. This
analysis will be able to compare not only different parallel algorithms and algebraic solvers, but also discretizations,
as demonstrated in Section~\ref{sec:results}.
%-------------------;
%  TAS-description  ;
%-------------------;
\begin{figure}[t]
\centering
\subfloat{\includegraphics[width=0.95\textwidth]{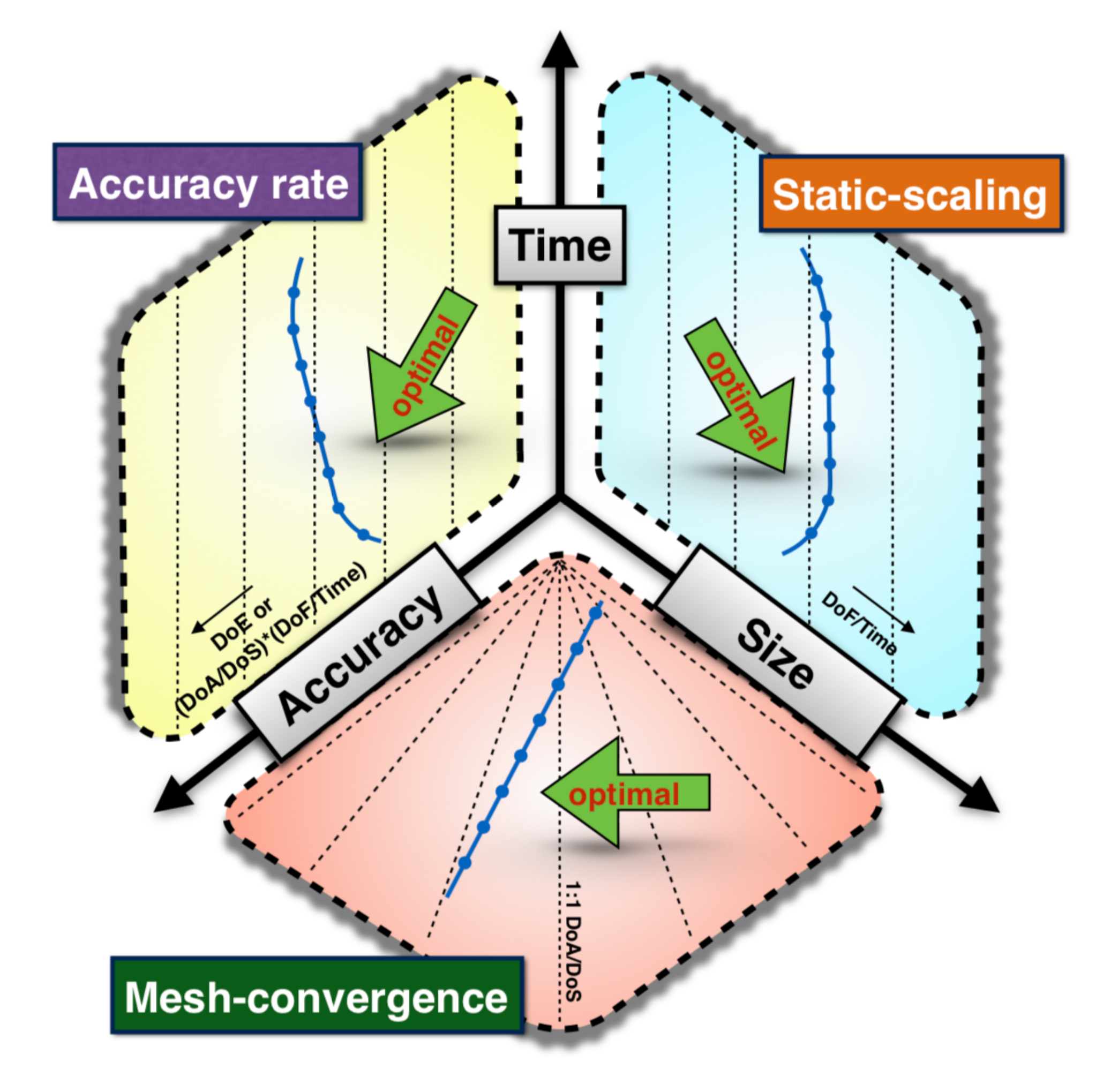}}
\caption{Pictorial description of the Time-Accuracy-Size (TAS) spectrum.\label{Fig:tasspectrum}}
\end{figure}

We now present the Time-Accuracy-Size (TAS) spectrum. In 
Figure~\ref{Fig:tasspectrum}, we show the relation of our new 
efficacy analysis to the existing mesh convergence and
static scaling plots. As outlined in 
\cite{ChangNakshatralaKnepleyJohnsson2017}, static-scaling
measures the degrees of freedom solved per second for a given 
parallelism. That is,
\begin{align}\label{eqn:dofpersec}
  \mbox{Static-scaling} \qquad\mathrm{measures}\qquad \left(\frac{\mathrm{size}/\mathrm{time}}{\mathrm{time}}\right)
\end{align}
We assume that the problems are of linear complexity, i.e.\ the time is in
$\mathcal{O}(N)$, so optimal scaling is indicated by a horizontal line 
as the problem size is increased. A higher computation rate indicates
that the algorithm matches the hardware well, but tells us little about
how accurate the solution is. Measures of the accuracy as a function of problem size,
however, are basic to numerical analysis, and usually referred to as mesh convergence,
\begin{align}\label{eqn:doaperdos}
  \mbox{Mesh-convergence} \qquad\mathrm{measures}\qquad \left(\frac{1/\mathrm{error}}{\mathrm{size}}\right)
\end{align}
where we use the inverse of error since we usually measure the negative logarithm of the error. Multiplying equations
\eqref{eqn:dofpersec} and \eqref{eqn:doaperdos} together, we arrive at rate for accuracy production,
\begin{align}\label{eqn:doapertime}
  \mbox{Accuracy rate} \qquad\mathrm{measures}\qquad
  \left(\frac{1/\mathrm{error}}{\mathrm{size}}\right)\times\left(\frac{\mathrm{size}/\mathrm{time}}{\mathrm{time}}\right) = \frac{1/(\mathrm{error}\times \mathrm{time})}{\mathrm{time}},
\end{align}
which is exactly our efficacy measure. An alternate derivation would be to scale the DoF count used in the
typical static-scaling analysis by the mesh convergence ratio, which we call \emph{true} static-scaling. This produces
the same measure, but slightly different scaling when logarithms are applied. Looking at equations~\eqref{eqn:dofpersec},
\eqref{eqn:doaperdos}, and \eqref{eqn:doapertime} give us the TAS spectrum, as visually depicted in
Figure~\ref{Fig:tasspectrum}. This figure illustrates how the new efficacy analysis can be applied to the existing mesh
convergence and static scaling plots.

%-------------------;
%  Accuracy rates  ;
%-------------------;
\begin{figure}[t]
\centering
\subfloat{\includegraphics[width=0.4\textwidth]{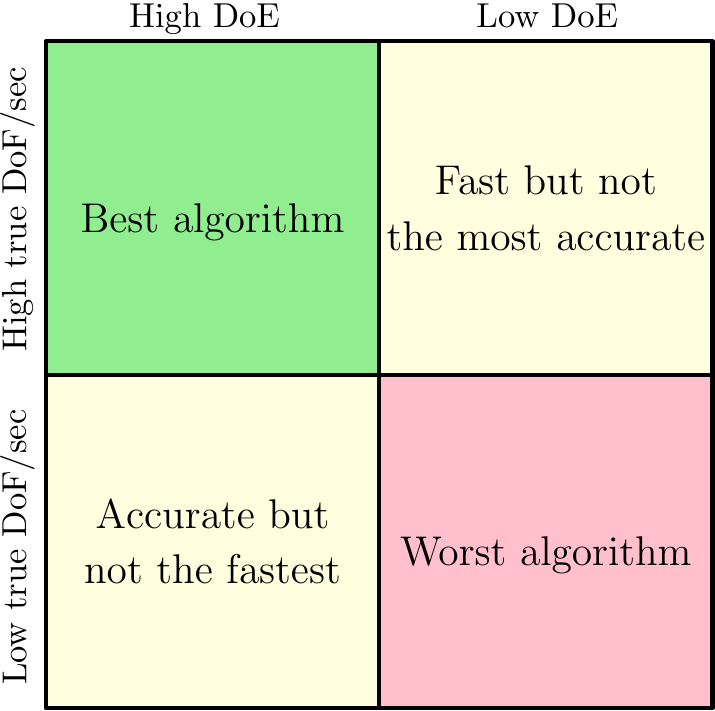}}
\caption{Interpretation of both the DoE and true DoF per second metrics of a 
particular algorithm. DoE is often an indicator of how accurate an algorithm is
for a given amount of time, whereas true DoF per time is often
an indicator of how fast an algorithm processes each DoF when each DoF is
scaled by how much accuracy it contributes.\label{Fig:doevstruedofsec}}
\end{figure}
\subsection{Interpreting the TAS spectrum}
The complete TAS spectrum could potentially have three or four different diagrams that provide a wealth of 
performance information. We now show the recommended order of interpretation of these diagrams as well as 
provide some guidelines on how to synthesize the data into an understandable
framework.
\begin{enumerate}
\item \textsf{Mesh convergence}: This diagram not only shows whether the actual $L_2$ convergence 
matches the predicted $\alpha$, but how much accuracy is attained for a given size. Any tailing off that occurs 
in the line plots could potentially be an issue of solver tolerance or implementation errors. Such tail-offs will 
drastically affect both accuracy rate diagrams, so this diagram could be an early warning sign for unexpected
behavior in those plots. Furthermore, the mesh convergence ratio (i.e., the DoA over DoS) can also be an early predictor
as to which discretizations or implementations will have better accuracy rates.
\item \textsf{Static-scaling}: This particular scaling analysis is particularly useful for examining both strong-scaling and
weak-scaling limits of parallel finite element simulations across various hardware architectures and 
software/solver implementations. Optimal scaling would produce a horizontal line or a ``sweet spot'' assuming that the
algorithm is of $\mathcal{O}(N)$ complexity. Any tailing off in these static-scaling plots will have a 
direct affect on the accuracy rate plots.
\item \textsf{DoE}: This metric gives the simplest interpretation of numerical accuracy and computational cost.
A high DoE is most desired, and if straight lines are observed in both the mesh convergence and static-scaling 
diagrams, the lines in this diagram should exhibit some predicted slope which will be discussed in the next section. 
Note that the size of the problem is not explicitly taken into account in these diagrams; these diagrams 
simply provide an easy visual on the ordering of the software implementation or finite element discretization.
\item \textsf{True static-scaling}: Optionally, the ordering shown in the DoE diagrams can be further verified 
through the true static-scaling plots. The information provided by this analysis simply tells us how fast the algorithm is being computed assuming that all DoF are given equal weighting. Figure \ref{Fig:doevstruedofsec} provides a simple guideline on how to simultaneously interpret both the DoE and true static-scaling diagrams. Note that the true static-scaling diagrams will not always produce a horizontal line, as the DoF is now scaled by the DoA over DoS ratio.
\end{enumerate}

\section{Theoretical Analysis}\label{sec:theory}
\begin{figure}[t]
  \centering
  \subfloat{\includegraphics[width=0.9\textwidth]{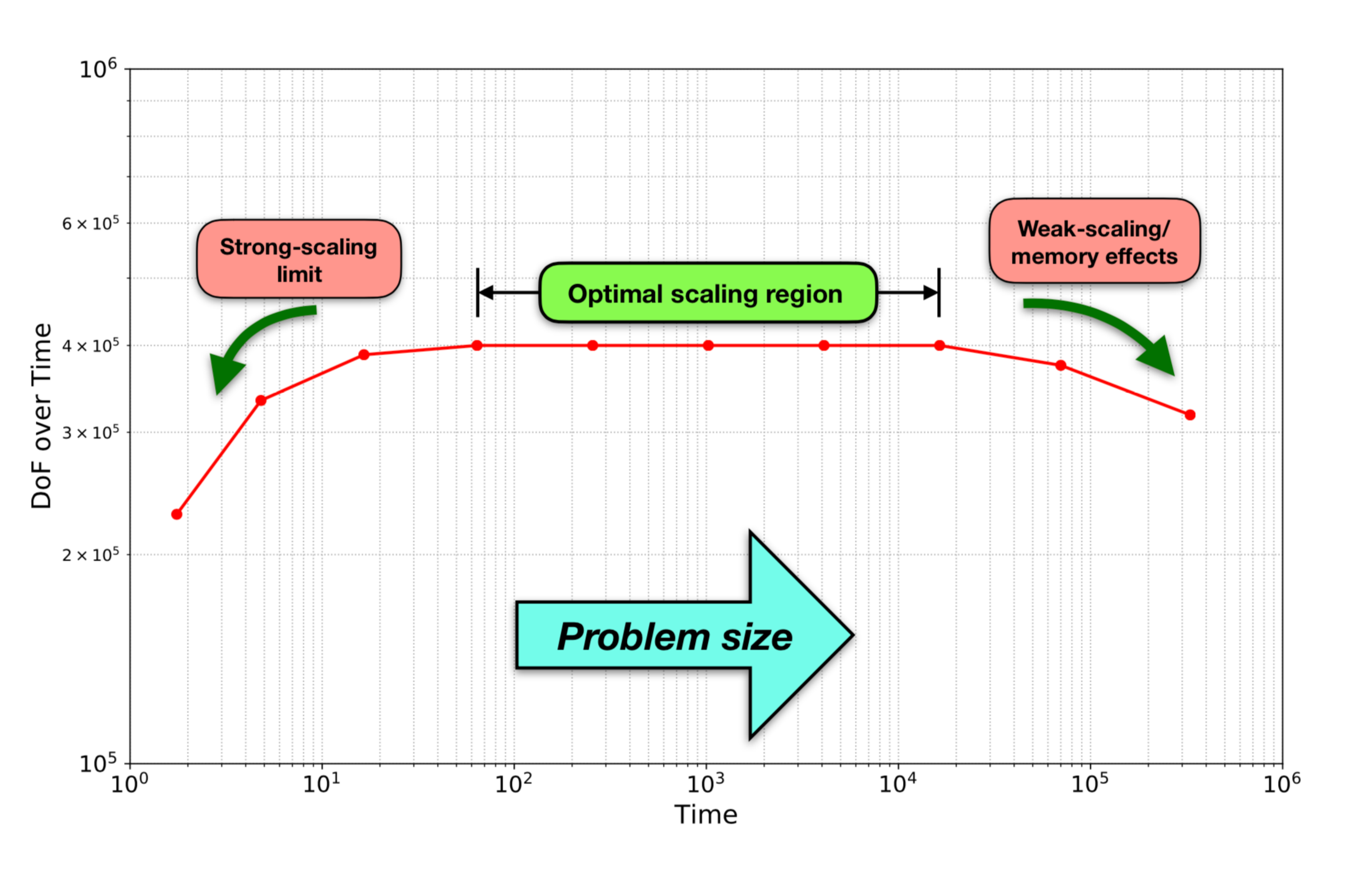}}
  \caption{Example static-scaling diagram and a description of the behavioral characteristics. This diagram
  is able to characterize both strong-scaling and weak-scaling effects across a variety of problem sizes.
  \label{Fig:examplestatic}}
\end{figure}
\begin{figure}[t]
  \centering
  \subfloat[$d = 2$]{\includegraphics[width=0.5\textwidth]{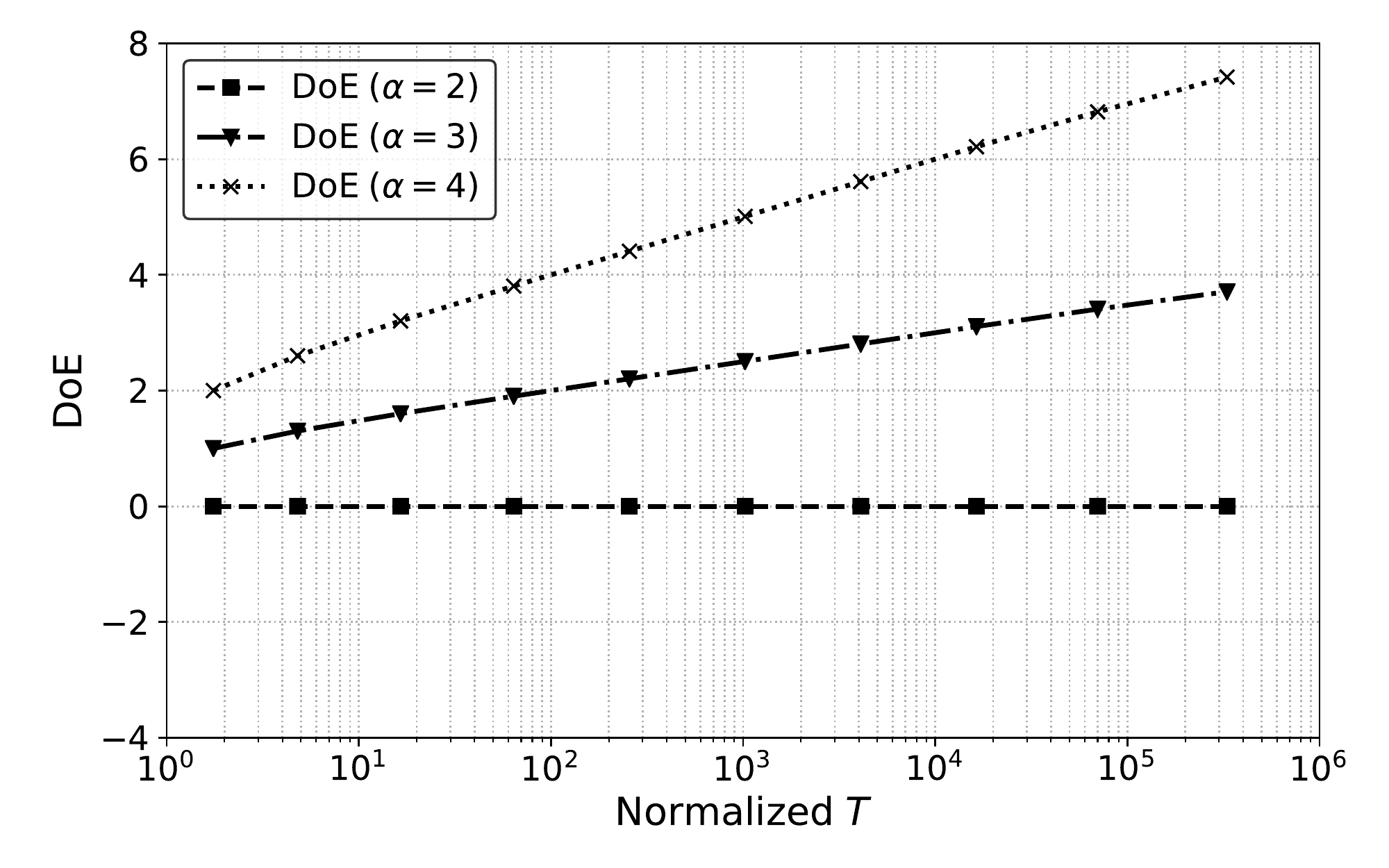}}
  \subfloat[$d = 3$]{\includegraphics[width=0.5\textwidth]{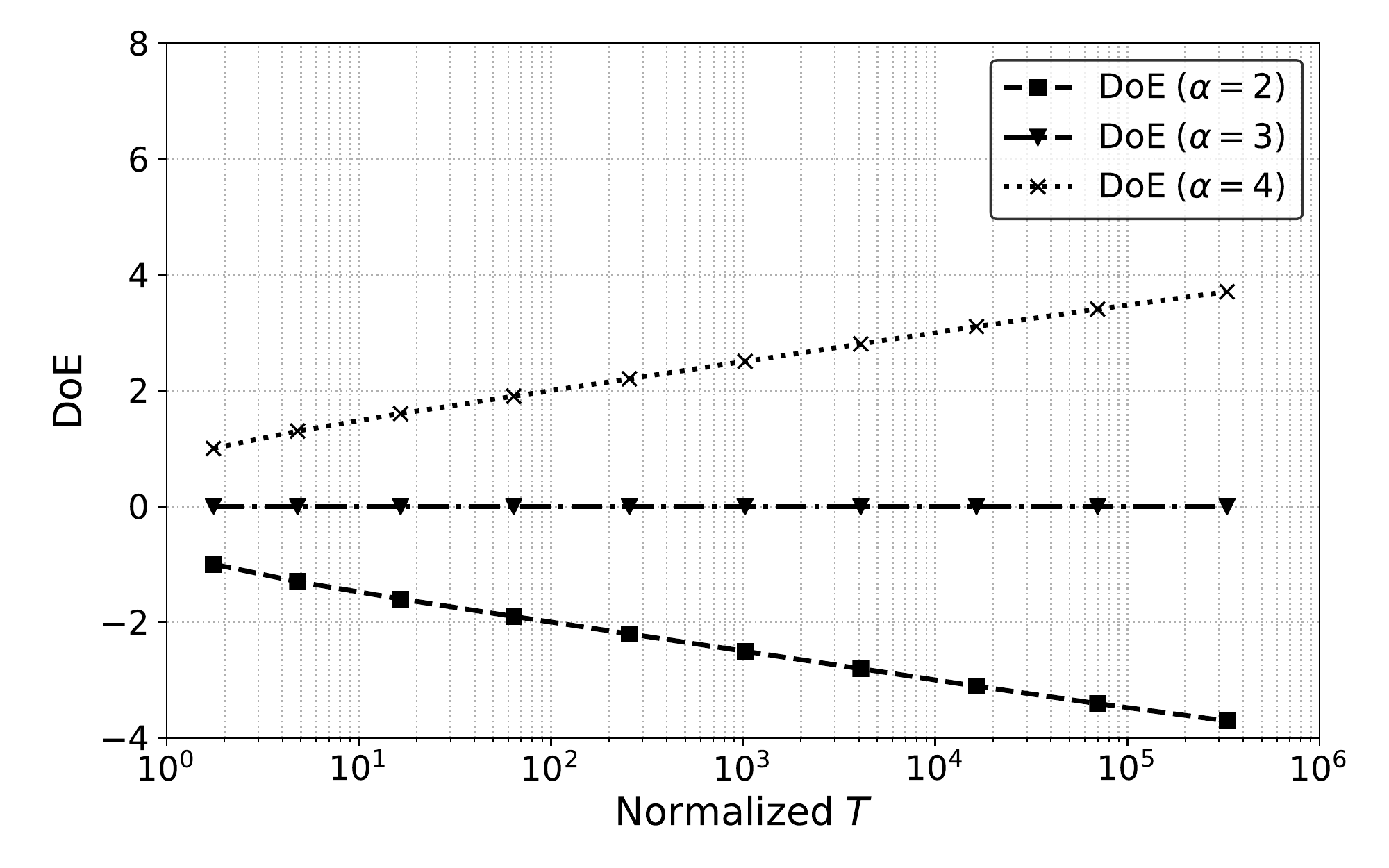}}
  \caption{Example DoE diagrams for 2D and 3D problems assuming $C = 10$, $W = 0.1$ and that
  $h$-sizes range from 1/10 to 1/5120. Both $\alpha$ and $d$ have a drastic effect on the slope of 
  these lines. \label{Fig:exampledoe}}
\end{figure}

Unlike the mesh convergence and static-scaling analyses, the accuracy 
rate diagrams, which consist of the DoE and the true DoF per time metrics, measure 
the accuracy achieved by a particular method in a 
given amount of time. In this Section, we will discuss the theoretical 
underpinning of these two diagrams and examine the behavior of the line plots.
The DoE is written as:
\begin{align}
\mathrm{DoE}\ = -\log_{10}(err\times T).
\end{align}
Recall that $err \le C h^\alpha$ is the $L_2$ norm of the error with a theoretical convergence
rate $\alpha$, which can be obtained directly using MMS, and $T$ is the time:
\begin{align}
  T = W h^{-d},
\end{align}
where $d$ is the spatial dimension, $h$ denotes the representative element length, and 
$C$ and $W$ are constants. Then for a given run, the digits of efficacy would be given by
\begin{align}
  \mathrm{DoE} &= -\log_{10}\left( C h^\alpha W h^{-d} \right) \\
               &= -\log_{10}\left( C W h^{\alpha - d} \right) \\
               &= (d - \alpha) \log_{10}(h) - \log_{10}(C W).
\end{align}
Since $C$ and $W$ are constants, the slopes of the DoE lines are only affected by $\alpha$ and $d$.
However, because of strong-scaling and weak-scaling limits, it is possible that the time $T$ may not always be
of linear complexity thus the slope may not actually be $d-\alpha$. Let us consider a simple static-scaling
example shown in Figure \ref{Fig:examplestatic}. It can be seen here that the DoF per time (or $N/T$) ratios 
indicate at what points both strong-scaling and weak-scaling/memory effects start to dominate for a given 
MPI parallelism. If $C = 10$, $W = 0.1$, and $h$-sizes ranging from 1/10 to 1/5120, we can see from 
Figure \ref{Fig:exampledoe} what type of slopes we could expect to see in the DoE diagrams assuming the
same $T$ from Figure \ref{Fig:examplestatic}. It can be seen that methods with higher order rates of convergence
are preferable as $h$ is refined. For this particular example's chosen parameters, it can also be 
seen that the strong-scaling effects skew a few of the data points at the beginning but the 
weak-scaling effects are nearly unseen. Such effects may not aways be negligible in these DoE diagrams but could be carefully noted from static-scaling.

\begin{figure}[t]
  \centering
  \subfloat[$h$-size range: 10$^{-1}$ to 10$^{-3}$]{\includegraphics[width=0.5\textwidth]{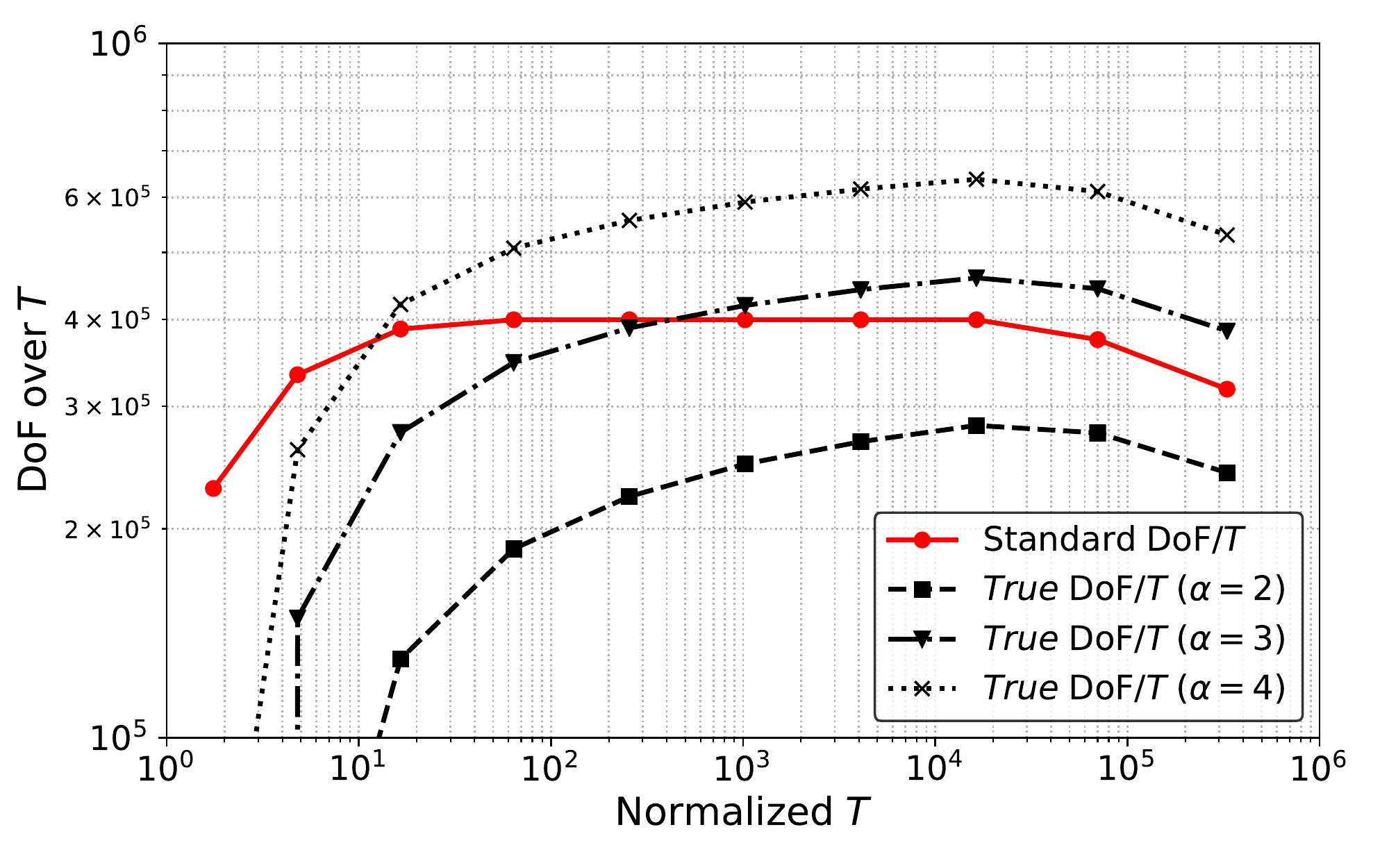}}
  \subfloat[$h$-size range: 10$^{-2}$ to 10$^{-4}$]{\includegraphics[width=0.5\textwidth]{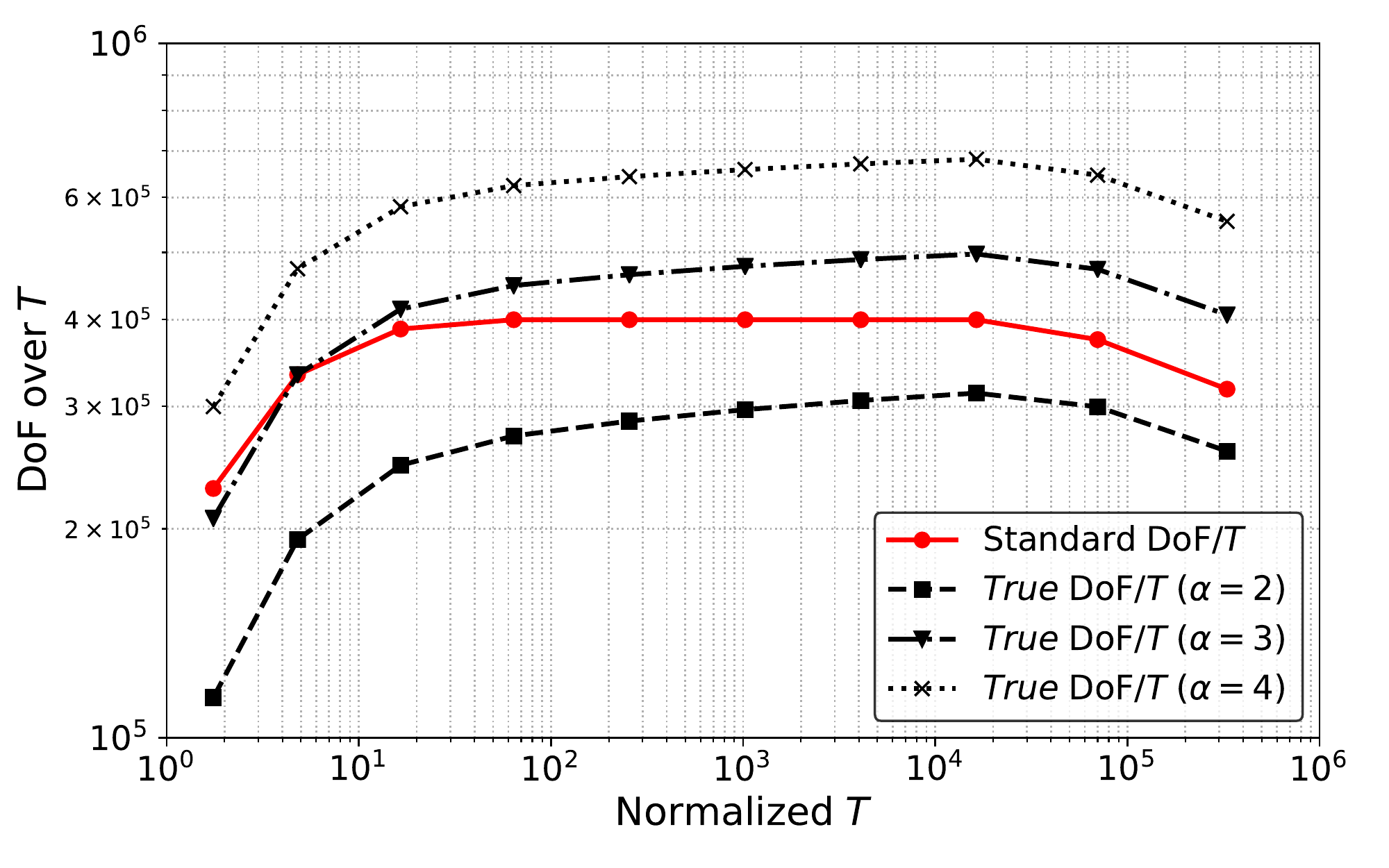}}\\
  \subfloat[$h$-size range: 10$^{-3}$ to 10$^{-5}$]{\includegraphics[width=0.5\textwidth]{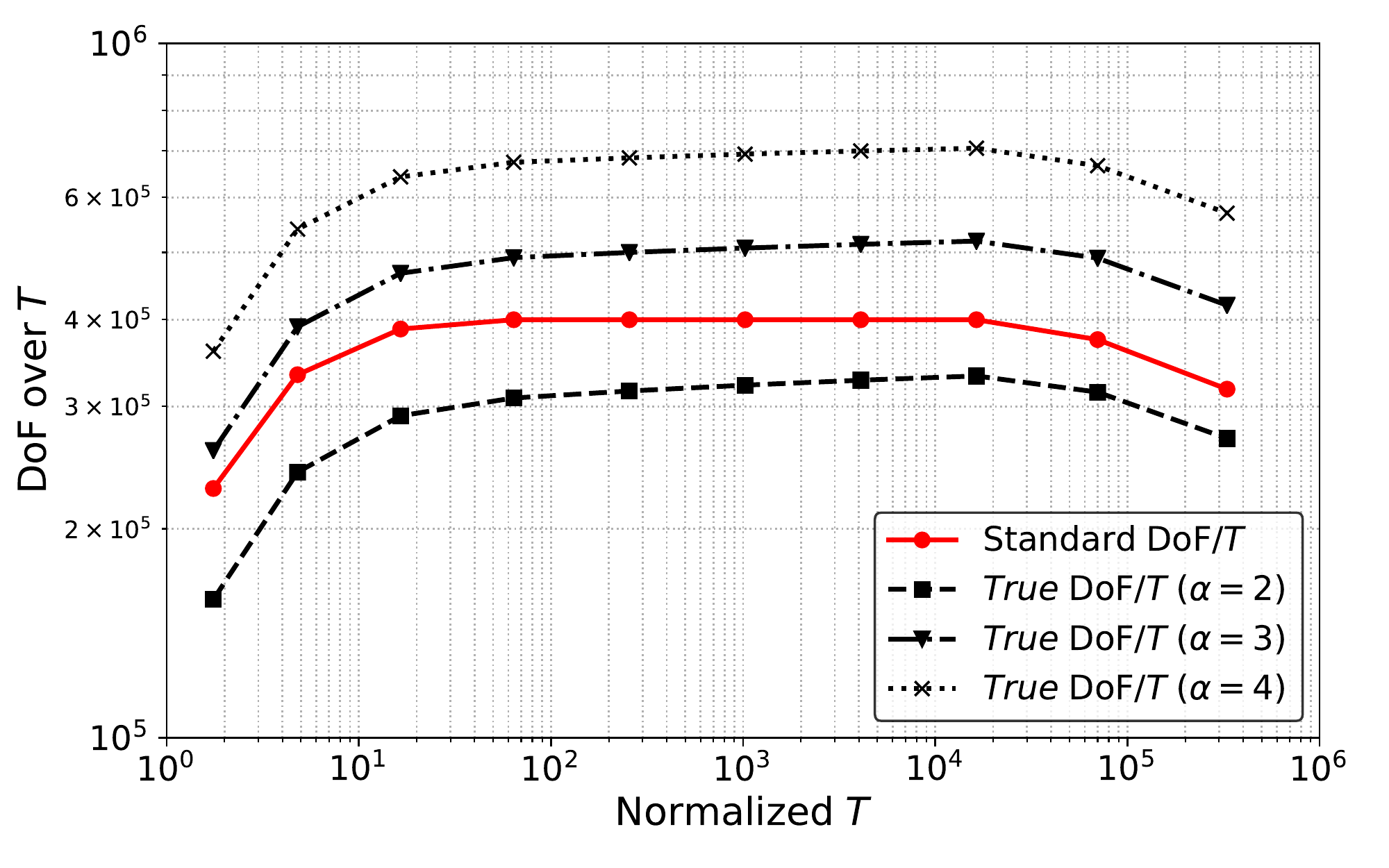}}
  \subfloat[$h$-size range: 10$^{-5}$ to 10$^{-7}$]{\includegraphics[width=0.5\textwidth]{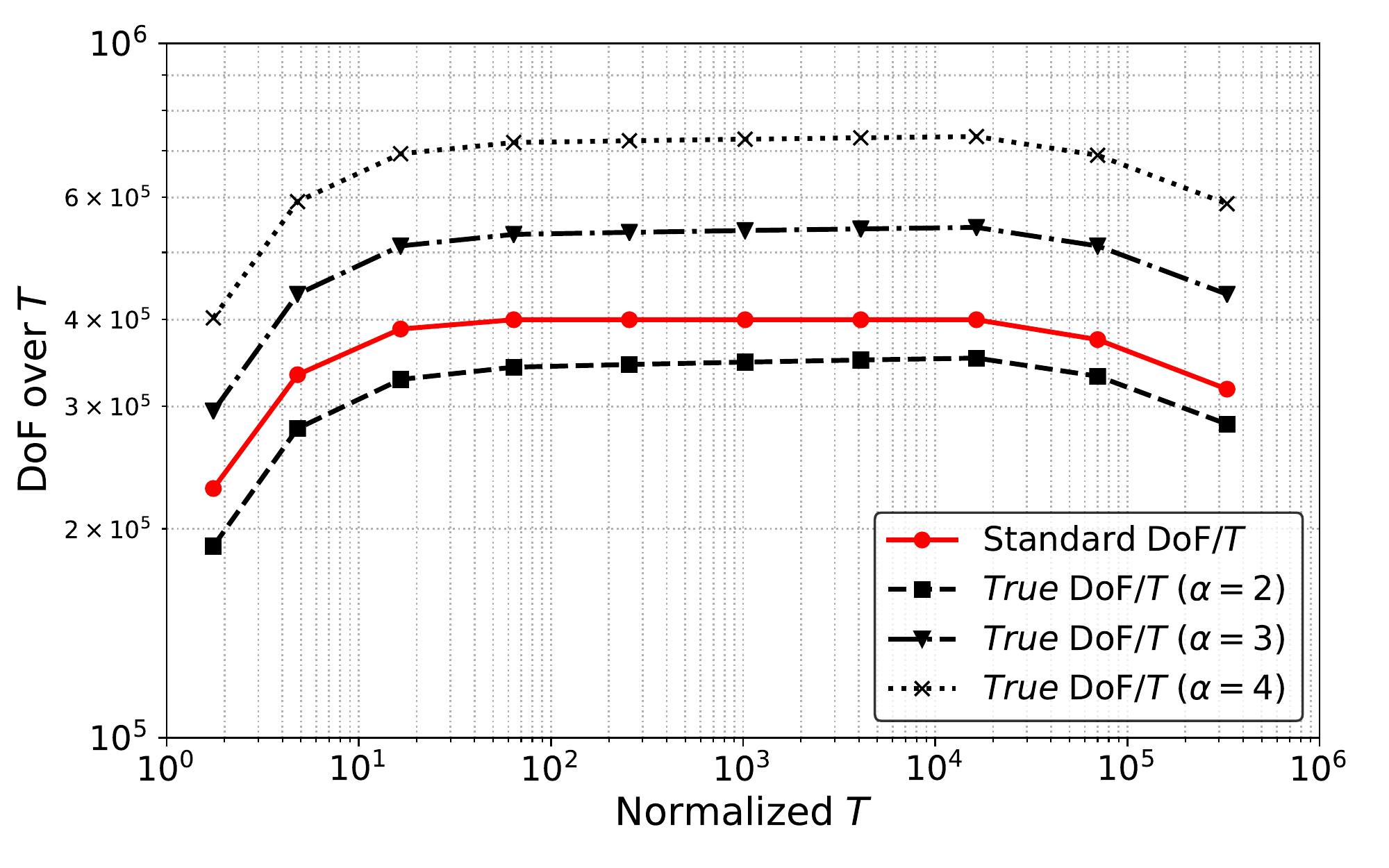}}\\
  \subfloat[$h$-size range: 10$^{-7}$ to 10$^{-9}$]{\includegraphics[width=0.5\textwidth]{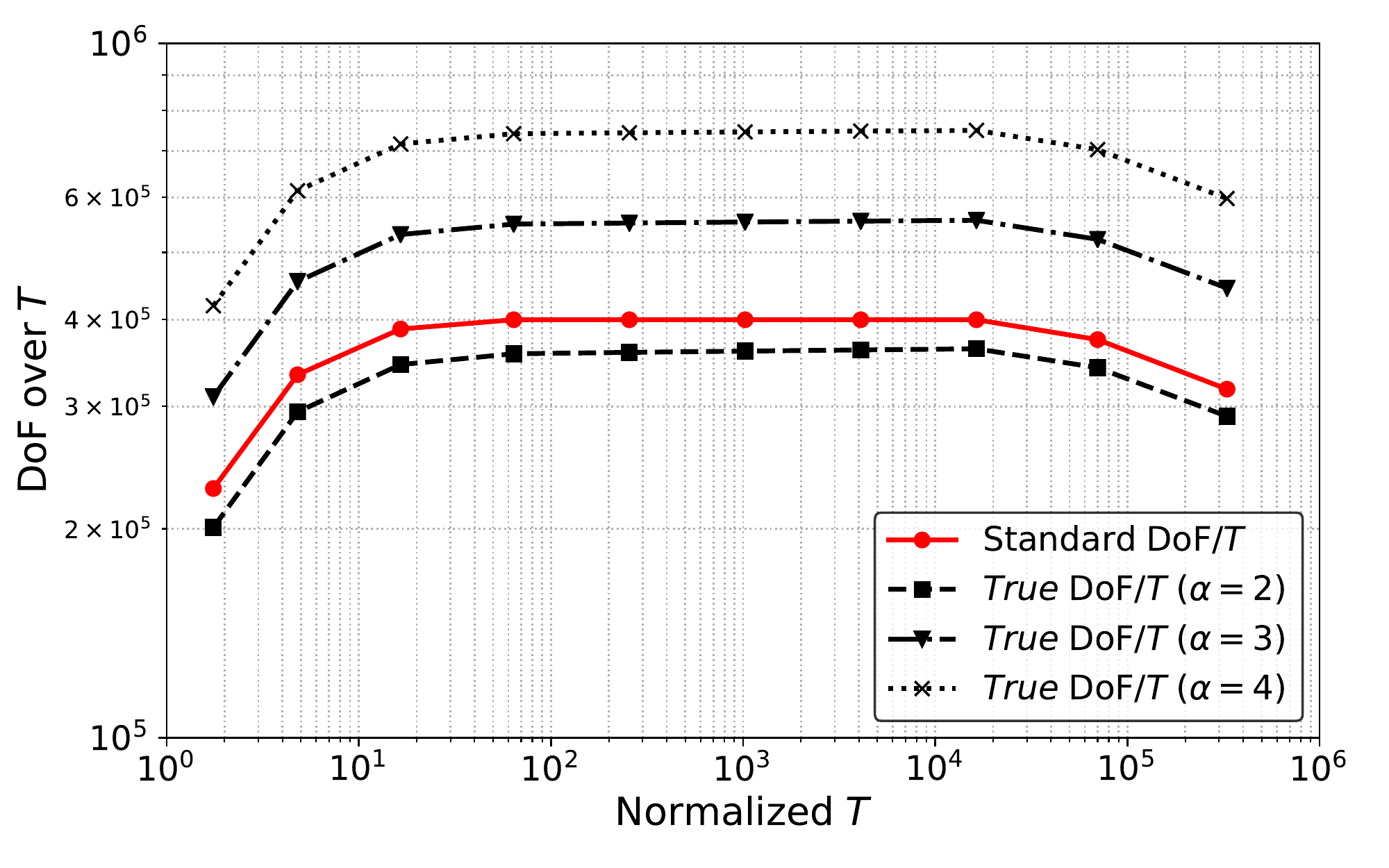}}
  \subfloat[$h$-size range: 10$^{-10}$ to 10$^{-12}$]{\includegraphics[width=0.5\textwidth]{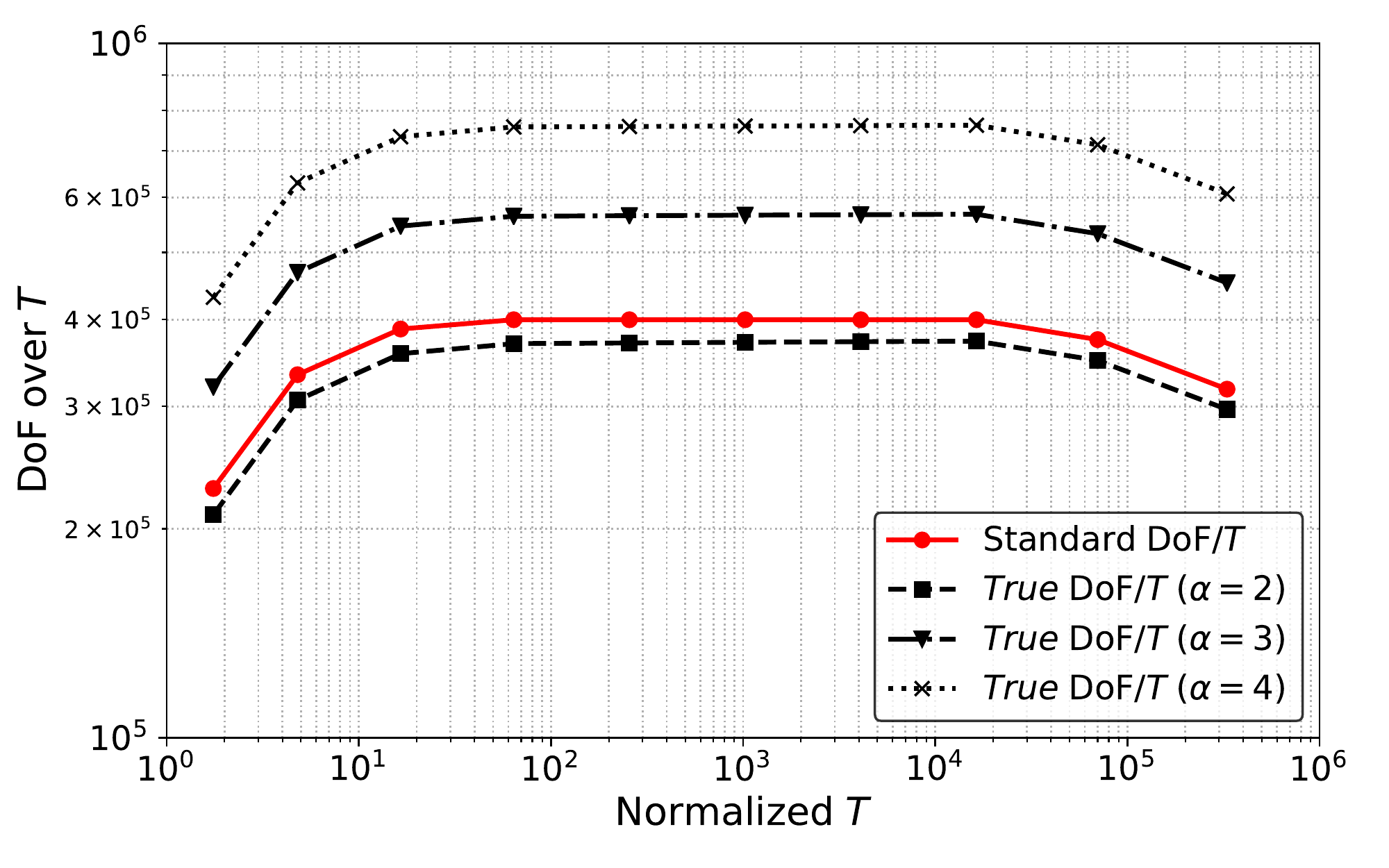}}
  \caption{True DoF per time rates in comparison to the original DoF per time rate for various $h$-sizes and $\alpha$. Let $d=2$, $C=10$, $D=4$, and the optimal DoF/$T = 4\times10^{5}$.\label{Fig:exampletruedof}}
\end{figure}
In true static-scaling, the DoF per time metric needs to be
scaled by the mesh convergence ratio DoA/DoS. Before we get into the
analysis of this scaling plot, we list a few key assumptions that must be
made in order for this to work:
\begin{enumerate}
\item Problem size DoF or $N = Dh^{-d}$ where $D$ is a constant. 
\item All computations are of linear complexity i.e., $\mathcal{O}(N)$.
\item Any tailing off from $\mathcal{O}(N)$ occurs due to  hardware related 
issues like latency from strong-scaling effects, memory bandwidth 
contention, or cache misses.
\item The problem size $N$ or DoF must be greater than 1.
\item $L_2$ error norm $err < 1.0$.
\end{enumerate}
Any violations to the last two assumptions would require that a different
logarithm rule be used or for the numbers to be scaled so that neither DoA nor DoS
are zero. The behavior of the true static-scaling line plot is given by:
\begin{align}
\left(\frac{\mathrm{DoA}}{\mathrm{DoS}}\right)\times\left(\frac{N}{T}\right) 
&= \left(\frac{-\log_{10}(C h^\alpha)}{\log_{10}(D h^{-d})}\right)\times\left(\frac{Dh^{-d}}{Wh^{-d}}\right) \\
&= \frac{-\log_{10}(C)-\log_{10}(h^\alpha)}{\log_{10}(D) + \log_{10}(h^{-d})}\times\left(\frac{D}{W}\right) \\
&= \frac{-\alpha\log_{10}(h) - \log_{10}(C)}{-d \log_{10}(h) + \log_{10}(D)}\times\left(\frac{D}{W}\right).
\end{align}
The variables $\alpha$, $d$, $C$, and $D$ will significantly impact the qualitative behavior of the true static-scaling
diagrams. As $h$ approaches zero, the $N/T$ ratio will slowly asymptote to a new $N/T$ ratio scaled by the factor
$\alpha/d$. Consider the following true-static scaling diagrams in Figure \ref{Fig:exampletruedof} when $d=2$, $C = 10$,
$D = 4$, and a variety of $h$-sizes are examined. It can be seen here that for larger $h$-sizes or coarser meshes, 
the DoF/$T$ lines are drastically skewed, and the optimal scaling regions are no longer horizontal. The 
relative ordering of the line plots in both accuracy rate diagrams depend significantly on the constants 
$C$, $D$, and $W$.

\section{Experimental Setup}\label{sec:setup}

In this paper, we only consider the Poisson equation,
\begin{align}\label{Eqn:model_problem}
  -\nabla^{2} u &= f,   &u &\in \Omega\\
  u &= u_0, &u &\in \Gamma_{D}
\end{align}
where $\Omega$ denotes the computational domain in $\mathbb{R}^d$, $\Gamma_D$ denotes its boundary, $u$ is the
scalar solution field, and $u_0$ are the prescribed Dirichlet boundary values.

The finite element discretizations considered are the Continuous Galerkin (CG) and Discontinuous Galerkin (DG)
methods. Various levels of both $h$- and $p$-refinement for the CG and DG methods are considered across different
software implementations. We do not consider other viable approaches such as the Hybridizable Discontinuous Galerkin
(HDG) method~\cite{kirby2012cg,FabienKnepleyRiviere2017} or mixed formulations~\cite{raviart1977mixed,
cockburn2009unified} but these will be addressed in future work. To this end, let us define $E$ as
an element belonging to a mesh $\mathcal{E}(\Omega)$. The relevant finite-dimensional function space for simplices is
\begin{align}
  \mathcal{U}_h &:= \left\{u_h \in L^{2}(\Omega):\; u_h\big|_E \in \mathcal{P}_{p}(E) \quad\forall\ E \in \mathcal{E}(\Omega)\right\},
\end{align}
and for tensor product cells is
\begin{align}
  \mathcal{U}_h &:= \left\{u_h \in L^{2}(\Omega):\; u_h\big|_E \in \mathcal{Q}_{p}(E) \quad\forall\ E \in \mathcal{E}(\Omega)\right\}.
\end{align}
Here $\mathcal{P}_{p}(E)$ denotes the space of polynomials in $d$ variables of degree less than or equal to $p$ over the
element $E$, and $\mathcal{Q}_{p}(E)$ is the space of $d$-dimensional tensor products of polynomials of degree less than
or equal to $p$. The general form of the weak formulation for equation \eqref{Eqn:model_problem} can be written as
follows: Find $u_h\in\mathcal{U}_h$ such that
\begin{align}\label{Eqn:weak_form_general}
  \mathcal{B}\left(v_h; u_h\right) = \mathcal{L}\left(v_h\right) \quad \forall\ v_h\in\mathcal{U}_h
\end{align}
where $\mathcal{B}$ and $\mathcal{L}$ denote the bilinear and linear forms, respectively.

\subsection{Finite element discretizations}

For the CG discretization, the solution $u_h$ is continuous at element boundaries, so that $\mathcal{U}_h$ is actually a
subspace of $H^1$, and the test functions satisfy $v_h = u_0$ on $\Gamma_D$. To present the DG formulation
employed in the paper, we introduce some notation. The boundary of a cell $E_i$ is denoted by $\partial E_i$. The
interior face between $E_i$ and $E_j$ is denoted by $\Gamma_{ij}$. That is,
\begin{align}
  \Gamma_{ij} = \partial E_i \cap \partial E_j
\end{align}
The set of all points on the interior faces is denoted by $\Gamma_{\mathrm{int}}$. Mathematically,
\begin{align}
  \Gamma_{\mathrm{int}} = \bigcup_{i,j}^{\mathcal{E}(\Omega)} \Gamma_{ij}
\end{align}
For an interior face, we denote the subdomains shared by this face by $E^{+}$ and $E^{-}$. The outward normals on this
face for these cells are, respectively, denoted by $\widehat{\mathbf{n}}^{+}$ and $\widehat{\mathbf{n}}^{-}$.
Employing Brezzi's notation~\cite{ArnoldBrezziCockburnMarini2002}, the average and jump operators on an interior face
are defined as follows
\begin{align}
  \big\{c\big\} := \frac{c^+ + c^-}{2} \quad \mathrm{and} \quad
  \big[\!\big[c\big]\!\big] := c^+ \widehat{\mathbf{n}}^+ + c^- \widehat{\mathbf{n}}^-
\end{align}
where
\begin{align}
  c^+ = c\vert_{\partial E^+} \quad \mathrm{and} \quad c^- = c\vert_{\partial E^-}
\end{align}
Let $  \Gamma_D$ denote the set of all boundary faces.  For a face $e\in \Gamma_D$, we then define $\{ c \} = c|_e  $, and $[\![c]\!] = c|_e \widehat{\mathbf{n}}_e.$  One of the most popular DG formulations is the \emph{Symmetric Interior Penalty} method, which for equation
\eqref{Eqn:weak_form_general} is written
%%Old formulation:
%\begin{align}
%  \mathcal{B}(v_h;u_h) &:= \Big(\nabla v_h;\;\nabla u_h\Big)_\Omega
%%<<<<<<< HEAD
%  - \Big(\big[\!\big[v_h\big]\!\big];\;\big\{\nabla u_h\big\}\Big)_{\Gamma_\mathrm{int}\cup\Gamma_D} - \Big(\big\{\nabla v_h\big\};\;\big[\!\big[u_h\big]\!\big]\Big)_{\Gamma_\mathrm{int}\cup\Gamma_D} \nonumber \\
%& + \alpha\Big(\big[\!\big[v_h\big]\!\big];\;\big[\!\big[u_h\big]\!\big]\Big)_{\Gamma_\mathrm{int}} + \gamma\Big(v_h;\;u_h\Big)_{\Gamma_D}\\
%  \mathcal{L}(v_h) &:= \Big(v_h;\;f \Big)_\Omega - \Big(\nabla v_h\cdot\widehat{\mathbf{n}}_{e};\;u_0\Big)_{\Gamma_D} 
%  + \gamma\Big(v_h;\;u_0\Big)_{\Gamma_D}
%%=======
%%  - \Big(\big[\!\big[v_h\big]\!\big];\;\big\{\nabla u_h\big\}\Big)_{\Gamma_\mathrm{int} \cup \Gamma_D} \nonumber \\
%% &- \Big(\big\{\nabla v_h\big\};\;\big[\!\big[u_h\big]\!\big]\Big)_{\Gamma_\mathrm{int} \cup \Gamma_D}
%%  +  \Big(\alpha \big[\!\big[v_h\big]\!\big];\;\big[\!\big[u_h\big]\!\big]\Big)_{\Gamma_\mathrm{int} \cup \Gamma_D}
%%  \\
%%  \mathcal{L}(v_h) &:= \Big(v_h;\;f \Big)_\Omega - \Big(\nabla v_h\cdot\widehat{\mathbf{n}}_{e};\;u_0\Big)_{  \Gamma_D} 
%%  -  \Big( \gamma v_h;\;u_0\Big)_{  \Gamma_D}
%%>>>>>>> 0589abeccd7986e125c15aa8d10f9e098c2df95b
%\end{align}
\begin{align}
  \mathcal{B}(v_h;u_h) 
  &:= 
  \sum_{E \in \mathcal{E}(\Omega) } \int_E \nabla v_h \cdot \nabla u_h 
  -
  \sum_{e \in \Gamma_\mathrm{int} \cup \Gamma_D} \int_e \{ v_h \} \cdot \big[\!\big[u_h\big]\!\big]
  -
  \sum_{e \in \Gamma_\mathrm{int} \cup \Gamma_D} \int_e \{ u_h \} \cdot \big[\!\big[v_h\big]\!\big]
  \nonumber 
  \\ 
  &+
  \sigma
  \sum_{e \in \Gamma_\mathrm{int}  } \int_e \frac{|e|}{|E|} \big[\!\big[v_h\big]\!\big]  \cdot \big[\!\big[u_h\big]\!\big]  
  + 
  \gamma
  \sum_{e \in \Gamma_D } \int_e \frac{|e|}{|E|}  v_h u_h
  \\
  \mathcal{L}(v_h) &:= \sum_{E \in \mathcal{E}(\Omega) } \int_E v_h f
  -
  \sum_{e \in \Gamma_\mathrm{D}  } \int_e u_0 \nabla v_h \cdot \widehat{\mathbf{n}}_{e}
  +
  \gamma
  \sum_{e \in \Gamma_\mathrm{D}  } \int_e \frac{|e|}{|E|} v_h u_0
\end{align}
where $\widehat{\mathbf{n}}_{e}$ denotes the outward normal on an exterior face, $|e|$ is the measure of a face in the given triangulation, $|E|$ is the measure of a cell in the given triangulation, and the penalty terms $\alpha$
and $\gamma$ are written as:
\begin{align}
\sigma &= \frac{(p + 1)(p + d)}{2d}\\
\gamma &= 2\alpha
\end{align}
as described in ~\cite{Shahbazi2005}.

%\begin{figure}[t]
%  \centering
%  \subfloat{\includegraphics[height=0.45\textheight]{meshes.pdf}}
%  \caption{The structured and unstructured grids considered in this paper.\label{Fig:meshes}}
%\end{figure}
\begin{figure}[t]
  \centering
  \subfloat{\includegraphics[width=0.32\textwidth]{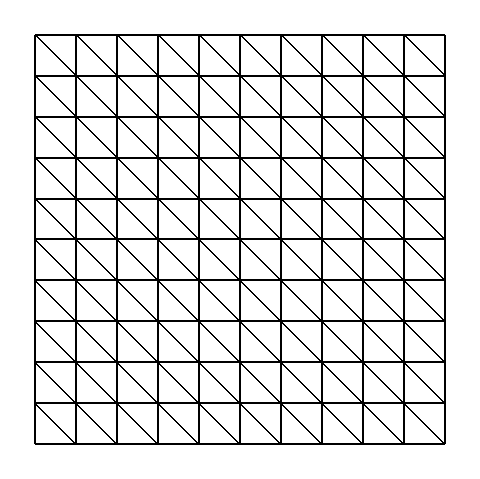}}
  \subfloat{\includegraphics[width=0.32\textwidth]{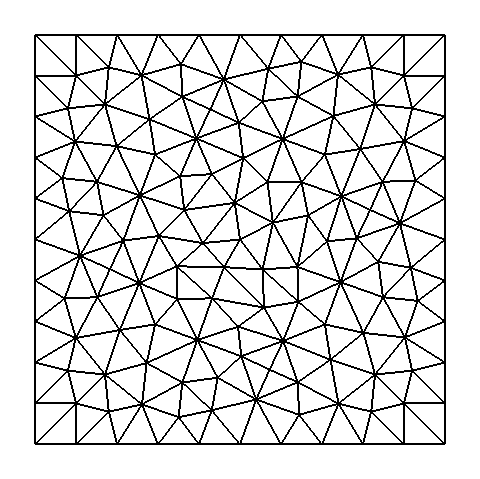}}
  \subfloat{\includegraphics[width=0.32\textwidth]{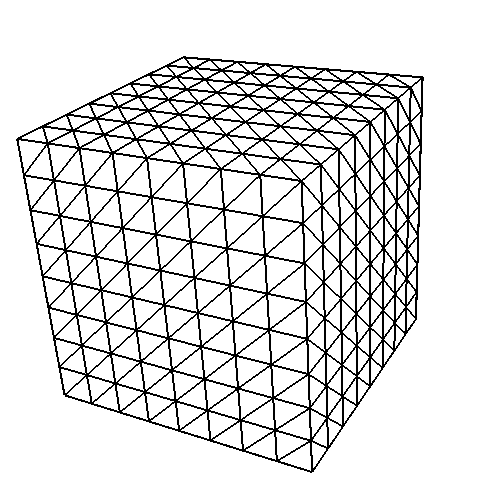}}\\
  \subfloat{\includegraphics[width=0.32\textwidth]{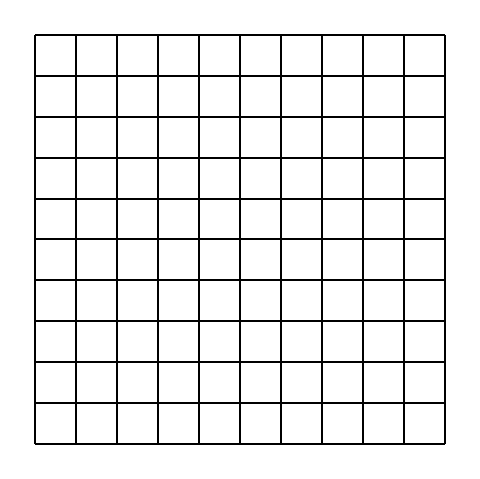}}
  \subfloat{\includegraphics[width=0.32\textwidth]{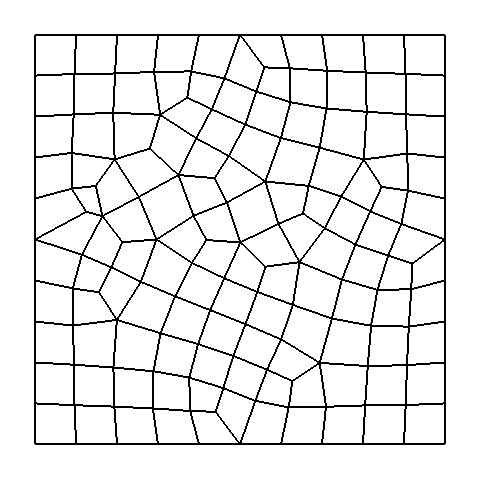}}
  \subfloat{\includegraphics[width=0.32\textwidth]{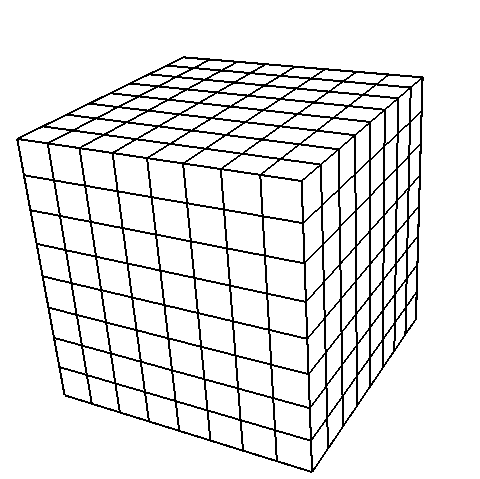}}
  \caption{The structured and unstructured grids considered in this paper.\label{Fig:meshes}}
\end{figure}
\subsection{Software and solver implementation}
In the next section, four different test problems are considered to demonstrate
the unique capabilities of the TAS spectrum analysis. For the first three problems,
the following sophisticated finite element software packages are examined: the 
C++ implementation of the FEniCS/Dolfin Project (Docker tag: 2017.1.0.r1), the 
the Python implementation of the Firedrake Project (SHA1: v0.13.0-1773-gc4b38f13),
and the C++ implementation of the deal.II library (Docker tag: v8.5.1-gcc-mpi-fulldepscandi-debugrelease). 
All three packages use various versions of PETSc for the solution of linear and nonlinear systems. 
The last problem utilizes the development version of PETSc (SHA1: v3.8.3-1636-gbcc3281268)
and its native finite element framework. The FEniCS/Dolfin/Firedrake software calculate 
the $L_2$ error norms by projecting the analytical solution $u$ onto a function space 3 degrees
order higher than the finite element solution $u_h$, whereas both deal.II and PETSc do the 
integral directly and use the same function space for both $u$ and $u_h$.

The test problems will be tested on the various meshes 
depicted in Figure~\ref{Fig:meshes}.  
The FEniCS/Dolfin and deal.II libraries use custom meshing, while 
Firedrake uses PETSc to manage unstructured meshes~
\cite{KnepleyKarpeev09,LangeMitchellKnepleyGorman2015,LangeKnepleyGorman2015},
and the hexahedral meshes are generated using the algorithms
described in~\cite{Bercea2016,McRae2016,Homolya2016}.
All timings will simply measure the finite element assembly 
and solve steps but not the mesh generation or any other 
preprocessing steps. 

The first three test problems presented in this paper will be 
solved using the conjugate gradient 
method with HYPRE's algebraic multigrid solver, BoomerAMG~\cite{falgout2002hypre}, and have 
a relative convergence criterion of $10^{-7}$. The last
problem will also consider two other multigrid solvers---the PETSc-native GAMG
preconditioner~\cite{adams2002evaluation,Adams-04} and Trilinos' Multi-Level (ML) solver~\cite{ml_users_guide}---but have a relative
convergence criterion of $10^{-9}$.

The first three tests are conducted on a single 3.5 GHz 
Quad-core Intel Core i5-7600 processor with 64 GB of 2400 MHz DDR4 memory.
The two (2D) problems are run in serial whereas the third (3D) problem is run across 4 MPI processes. 
The last problem is conducted on the Cori Cray XC40 system at the National Energy Research Scientific Computing Center (NERSC), 
and utilizes 32 Intel Xeon E5-2698v3 (``Haswell'') nodes for a total of 1024 MPI processes.
% RTM: I removed the registered trademark symbols above. Almost all journal style guidelines specify to not use them.
% Plus, we're not Intel and thus have no legal need to use them.

% composed solvers BruneKnepleySmithTu15, bkmms2012
\section{Computational Results}\label{sec:results}
%----------------------;
%  Subsection: Test 1  ;
%----------------------;
\subsection{Test \#1: Software implementation and mesh types}
%--------------------;
%  Table: Test 1  ;
%--------------------;
{\tiny \begin{table}[b]
\center
\caption{Comparison of CG1 across packages\label{Tab:test1}}
\begin{tabular}{ll|cc|cc|cc|cc}
\hline
\multirow{2}{*}{DoF} &\multirow{2}{*}{DoS} & 
\multicolumn{2}{c|}{FEniCS - triangles} & 
\multicolumn{2}{c|}{Firedrake - triangles} &
\multicolumn{2}{c|}{Firedrake - quads} &
\multicolumn{2}{c}{deal.II - quads} \\
& & DoA & DoA/DoS & DoA & DoA/DoS & DoA & DoA/DoS & DoA & DoA/DoS  \\
\hline
\multicolumn{10}{c}{Structured}\\
\hline
121 & 2.08 & 1.11 & 0.52 & 1.08 & 0.50 & 1.32 & 0.63 &  1.71 & 0.82  \\
441 & 2.64 & 1.66 & 0.61 & 1.65 & 0.61 & 1.91 & 0.72 &  2.31 & 0.87 \\
1681 & 3.23 & 2.25 & 0.68 & 2.24 & 0.68 & 2.50 & 0.78 &  2.92 & 0.90\\\
6561 & 3.82 & 2.84 & 0.73 & 2.84 & 0.73 & 3.10 & 0.81 &  3.52 & 0.92 \\
25921 & 4.41 & 3.41 & 0.77 & 3.44 & 0.77 & 3.71 & 0.84 &  4.12 & 0.93 \\
103041 & 5.01 & 4.04 & 0.79 & 4.04 & 0.79 & 4.31 & 0.86 & 4.72 & 0.94\\
\hline
\hline
\multicolumn{10}{c}{Unstructured}\\
\hline
142 & 2.15 & 1.27 & 0.59 & 1.27 & 0.59 & 1.30 & 0.63 &  1.63 & 0.76  \\
525 & 2.72 & 1.84 & 0.68 & 1.84 & 0.68 & 1.89 & 0.72 &  2.23 & 0.82  \\
2017 & 3.30 & 2.44 & 0.74 & 2.44 & 0.74 & 2.49 & 0.77 & 2.83 & 0.86  \\
7905 & 3.90 & 3.04 & 0.78 & 3.04 & 0.78 & 3.09 & 0.81 & 3.43 & 0.88  \\
31297 & 4.50 & 3.64 & 0.81 & 3.64 & 0.81 & 3.69 & 0.84 &4.03 & 0.90  \\
124545 & 5.10 & 4.24 & 0.83 & 4.24 & 0.83 & 4.29 & 0.86 & 4.64 & 0.91  \\
\hline
\end{tabular}
\end{table}}
%-------------------;
%  Figures: Test 1  ;
%-------------------;
\begin{figure}[t]
  \centering
  \subfloat[Mesh convergence: Structured]{\includegraphics[width=0.4\textwidth]{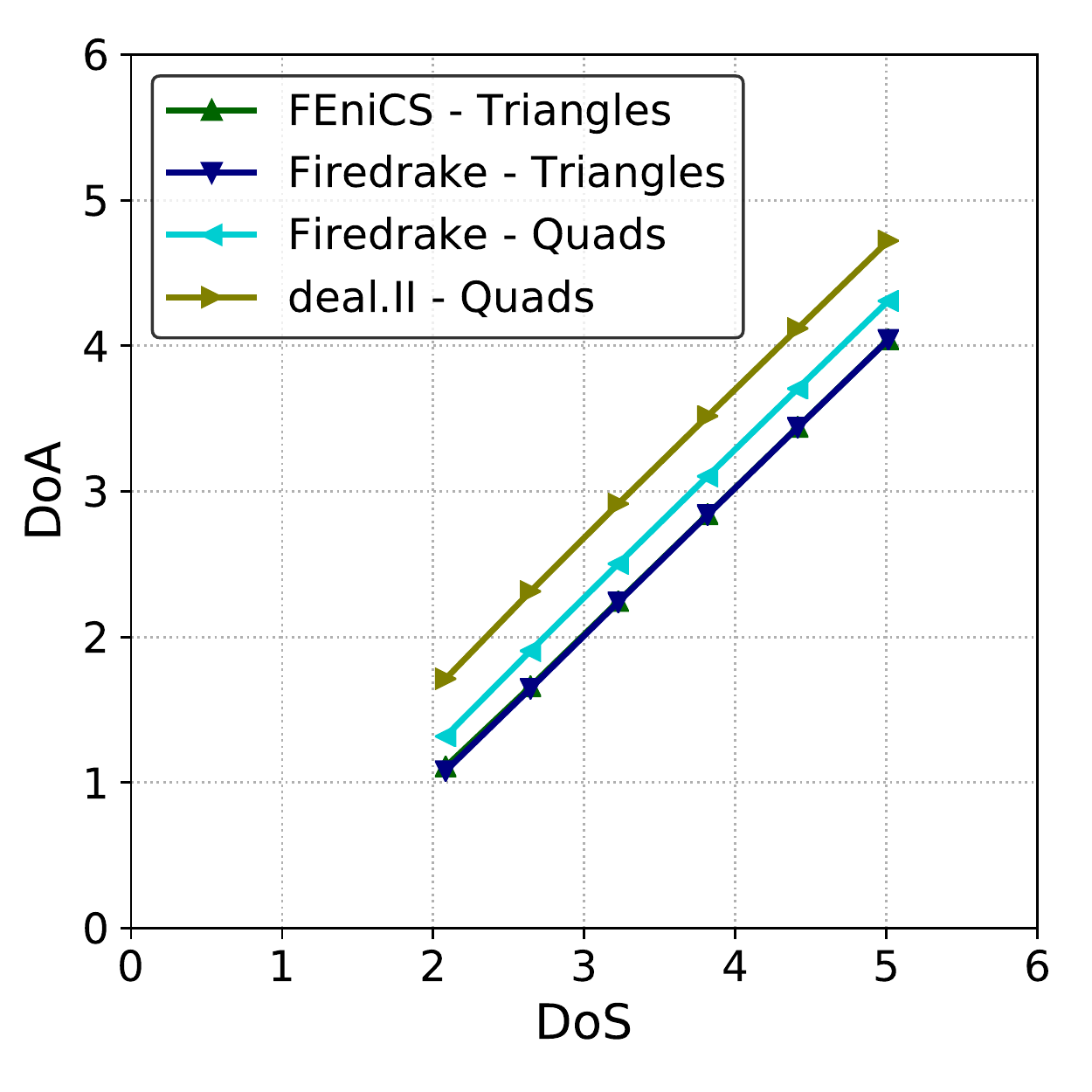}}
  \subfloat[Mesh convergence: Unstructured]{\includegraphics[width=0.4\textwidth]{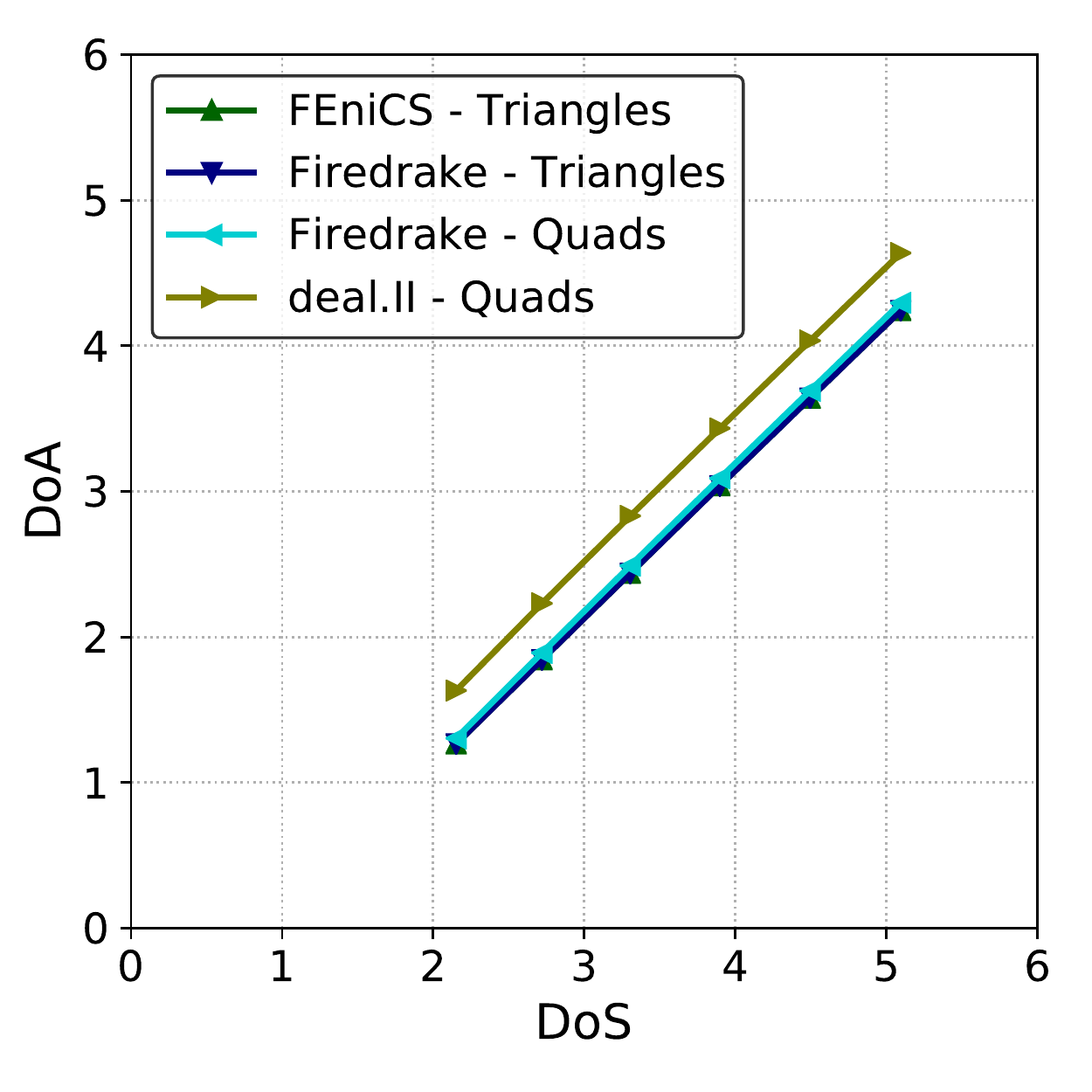}}\\
  \subfloat[Static-scaling: Structured]{\includegraphics[width=0.5\textwidth]{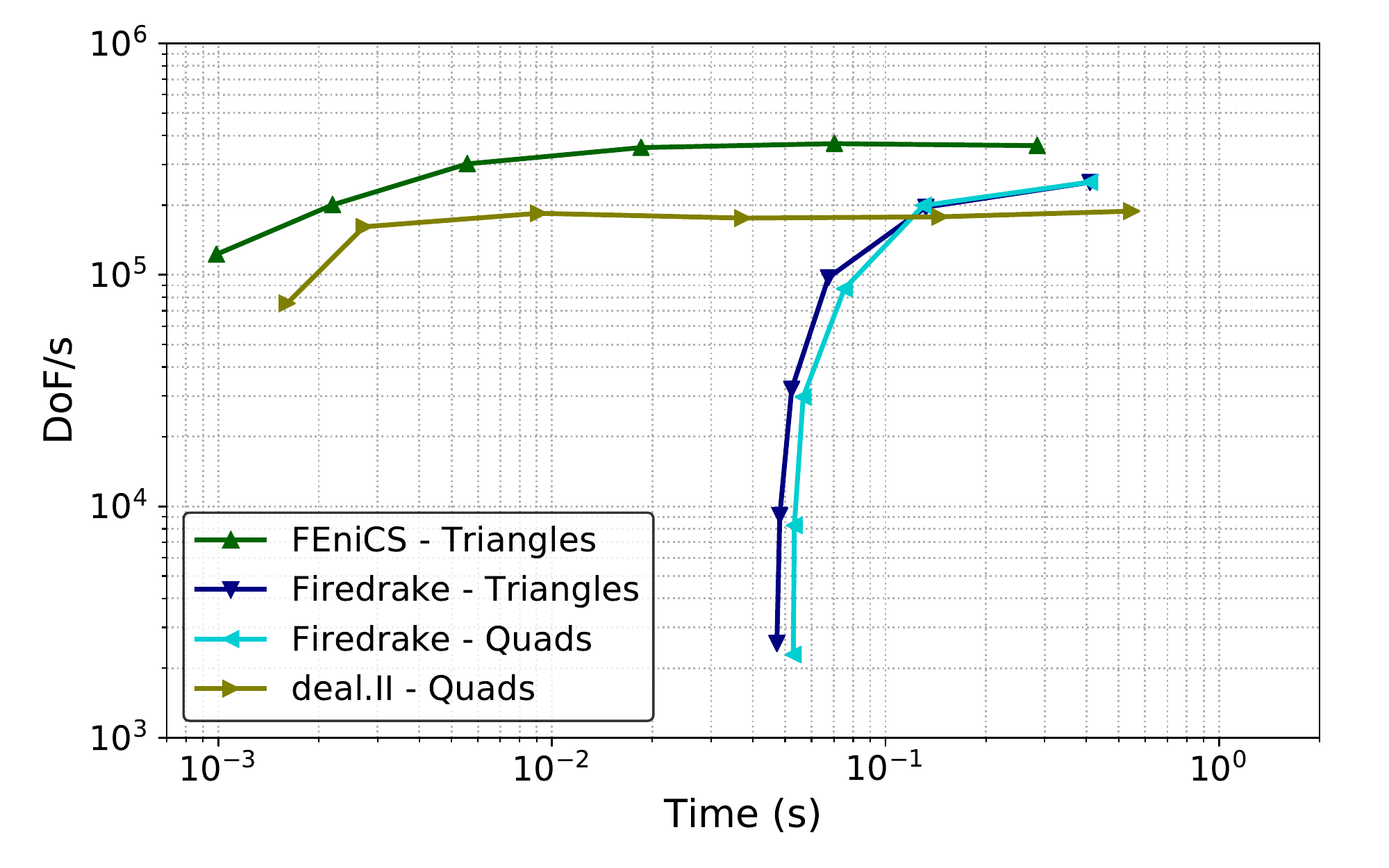}}
  \subfloat[Static-scaling: Unstructured]{\includegraphics[width=0.5\textwidth]{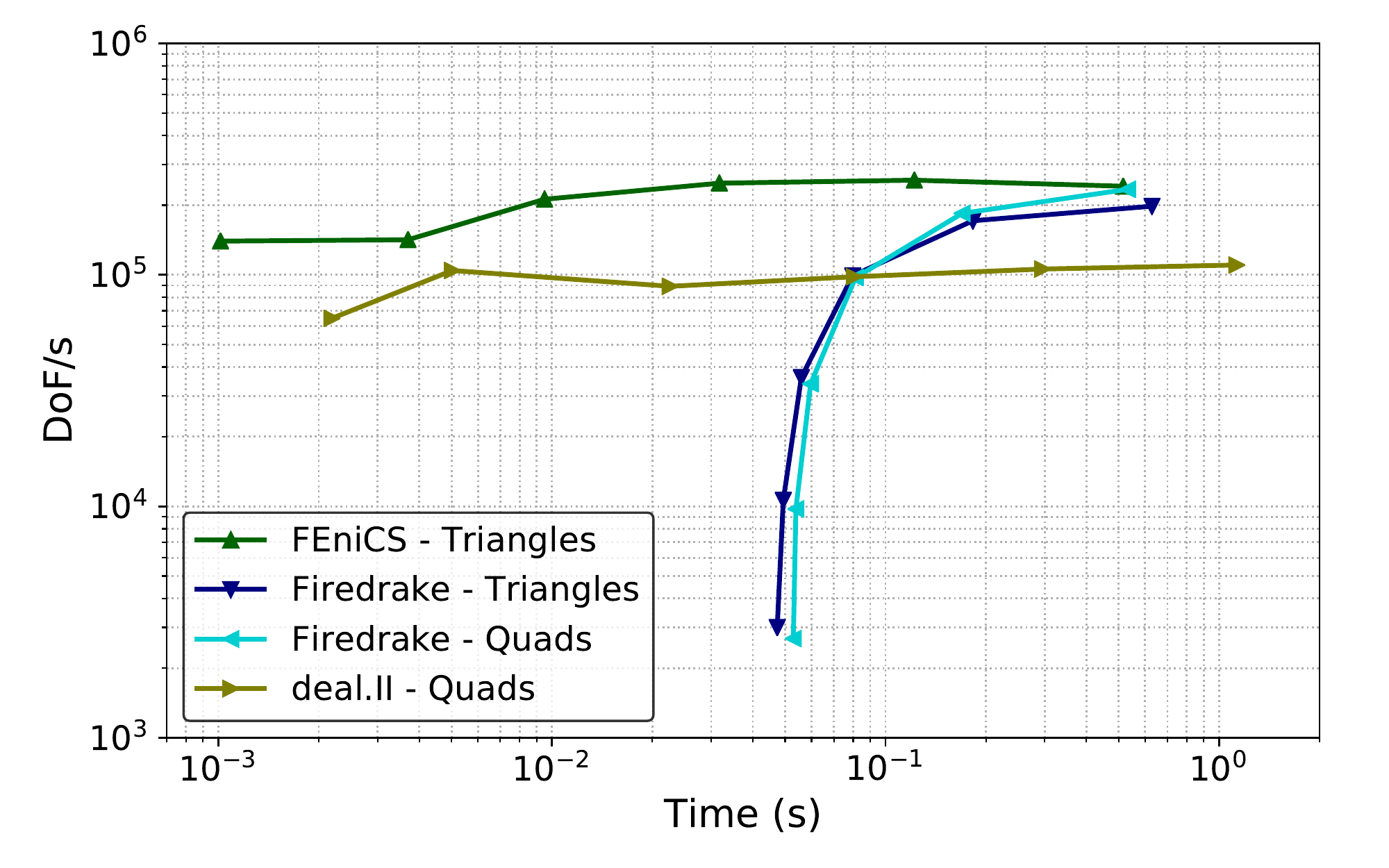}}
  \caption{Test \#1: Mesh convergence and static-scaling comparing various finite 
  element software packages on structured and unstructured 2D grids.\label{Fig:Test1meshstatic}}
\end{figure}
\begin{figure}[t]
  \centering
  \subfloat[DoE: Structured]{\includegraphics[width=0.5\textwidth]{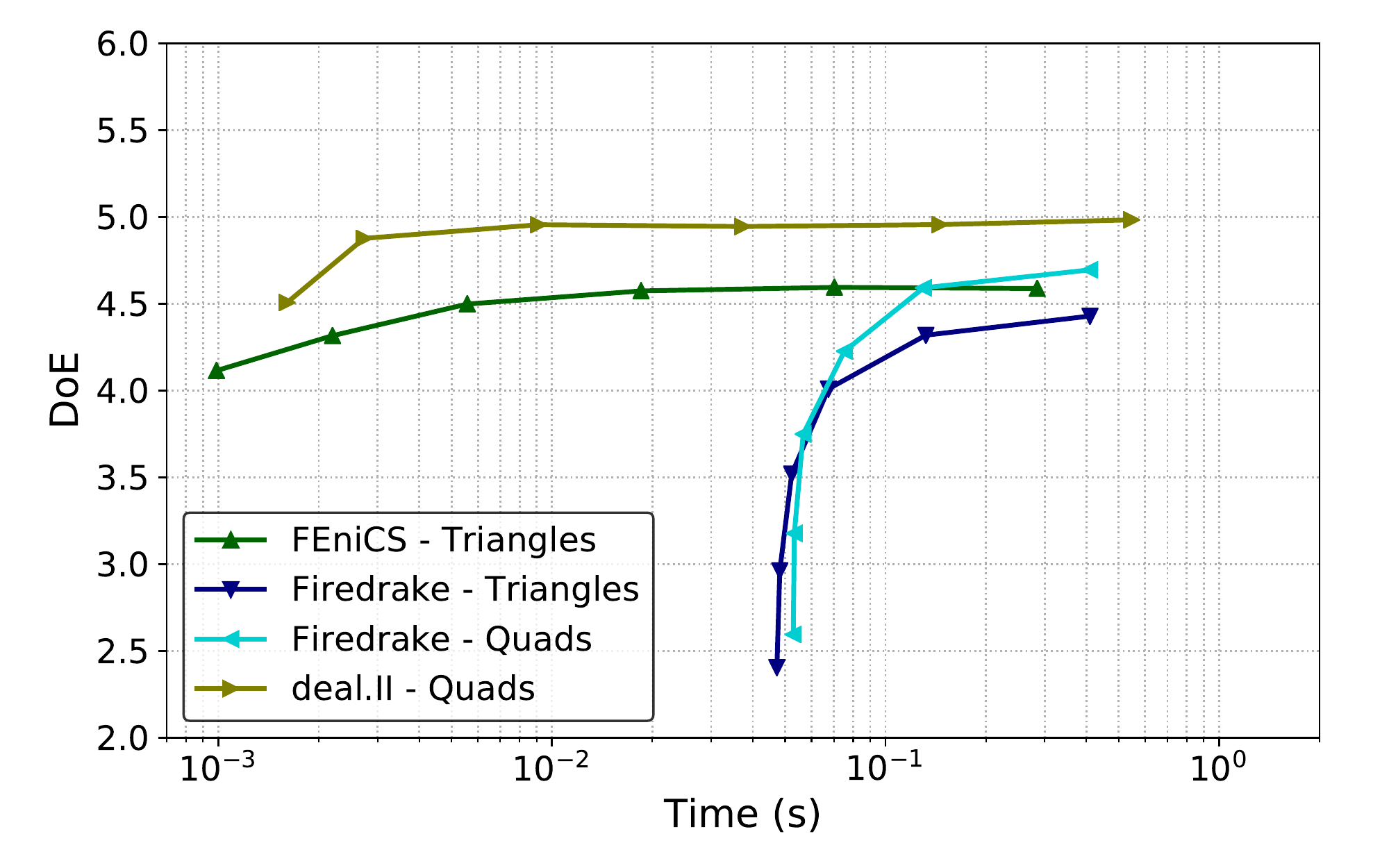}}
  \subfloat[DoE: Unstructured]{\includegraphics[width=0.5\textwidth]{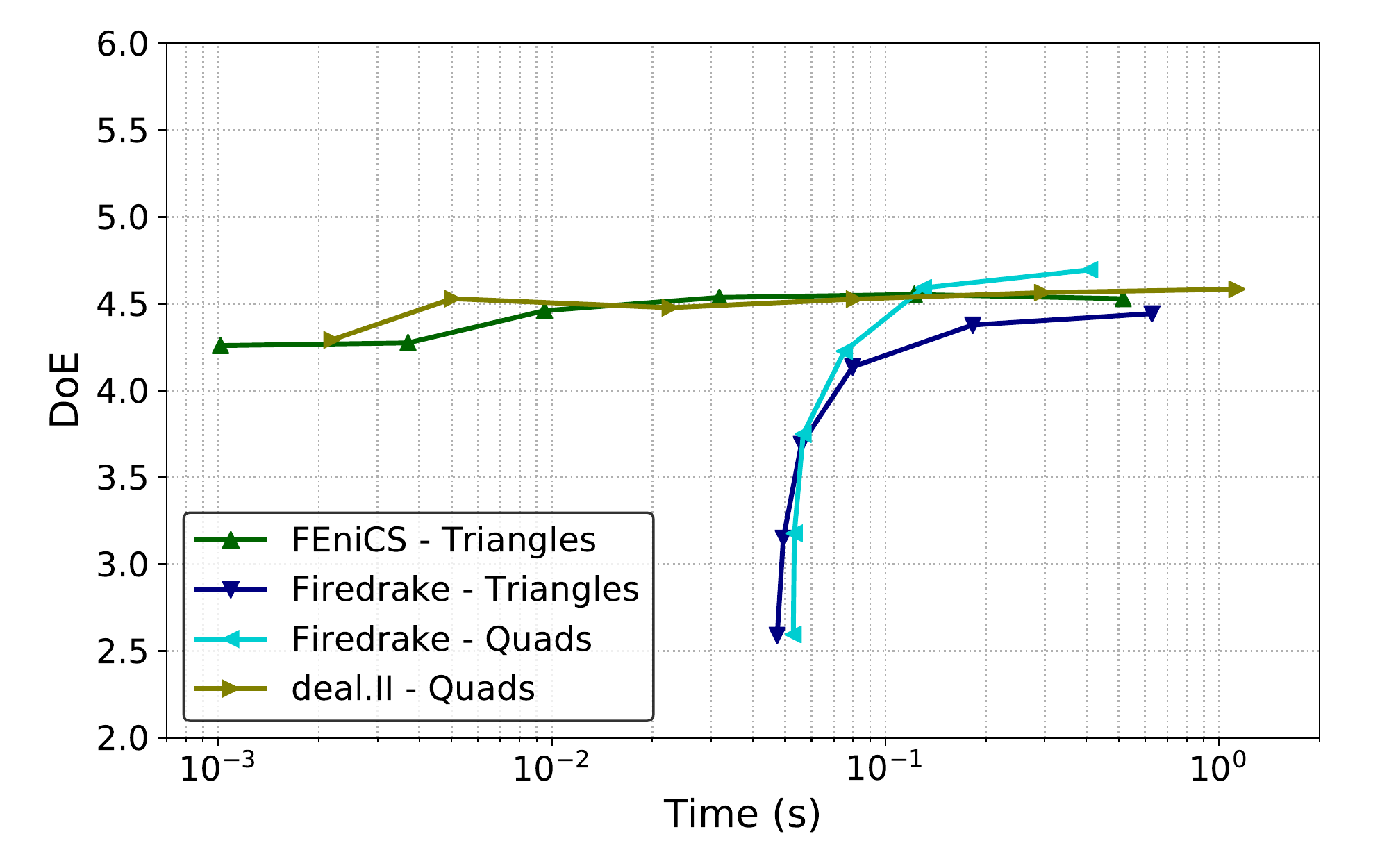}}\\
  \subfloat[True static-scaling: Structured]{\includegraphics[width=0.5\textwidth]{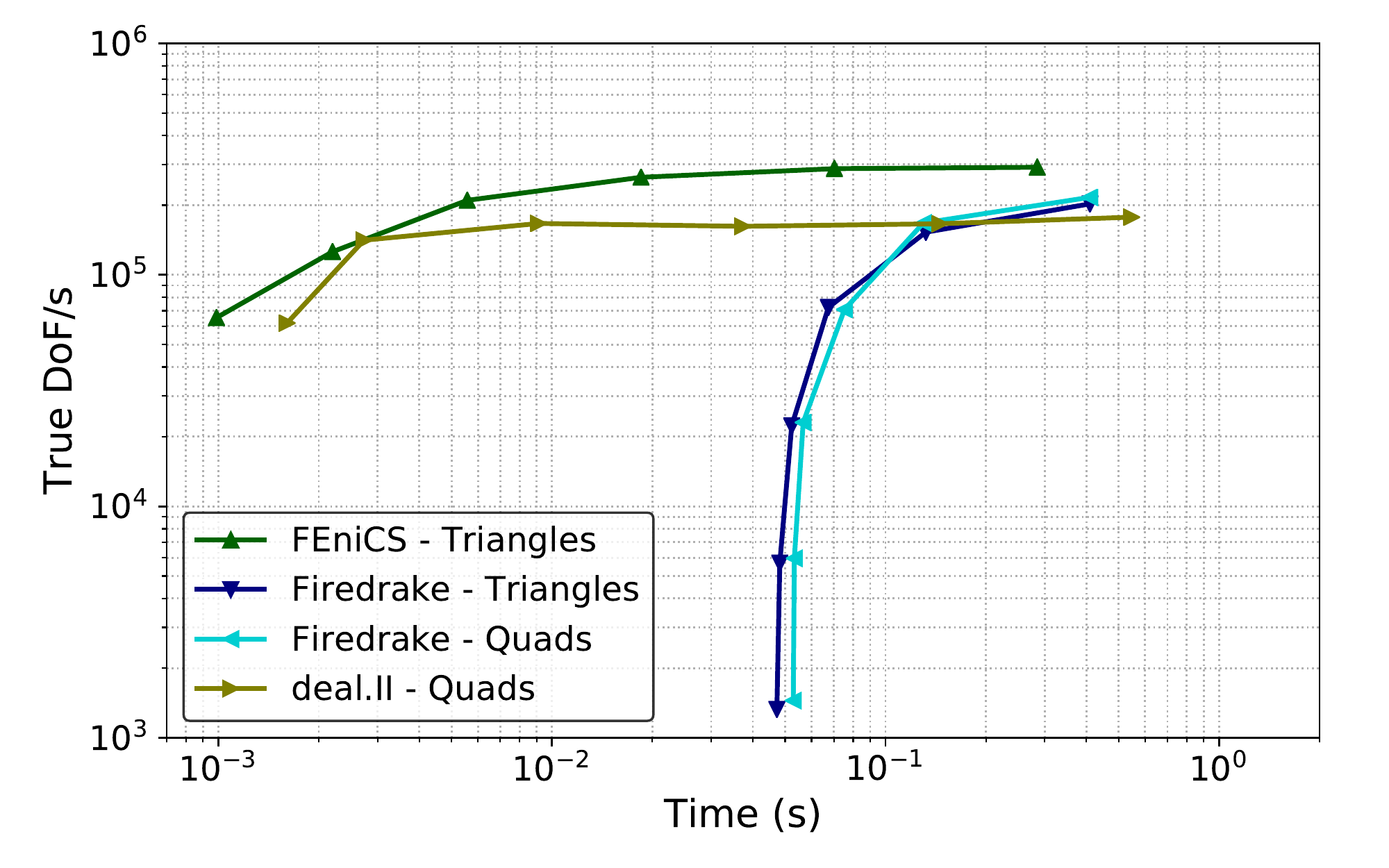}}
  \subfloat[True static-scaling: Unstructured]{\includegraphics[width=0.5\textwidth]{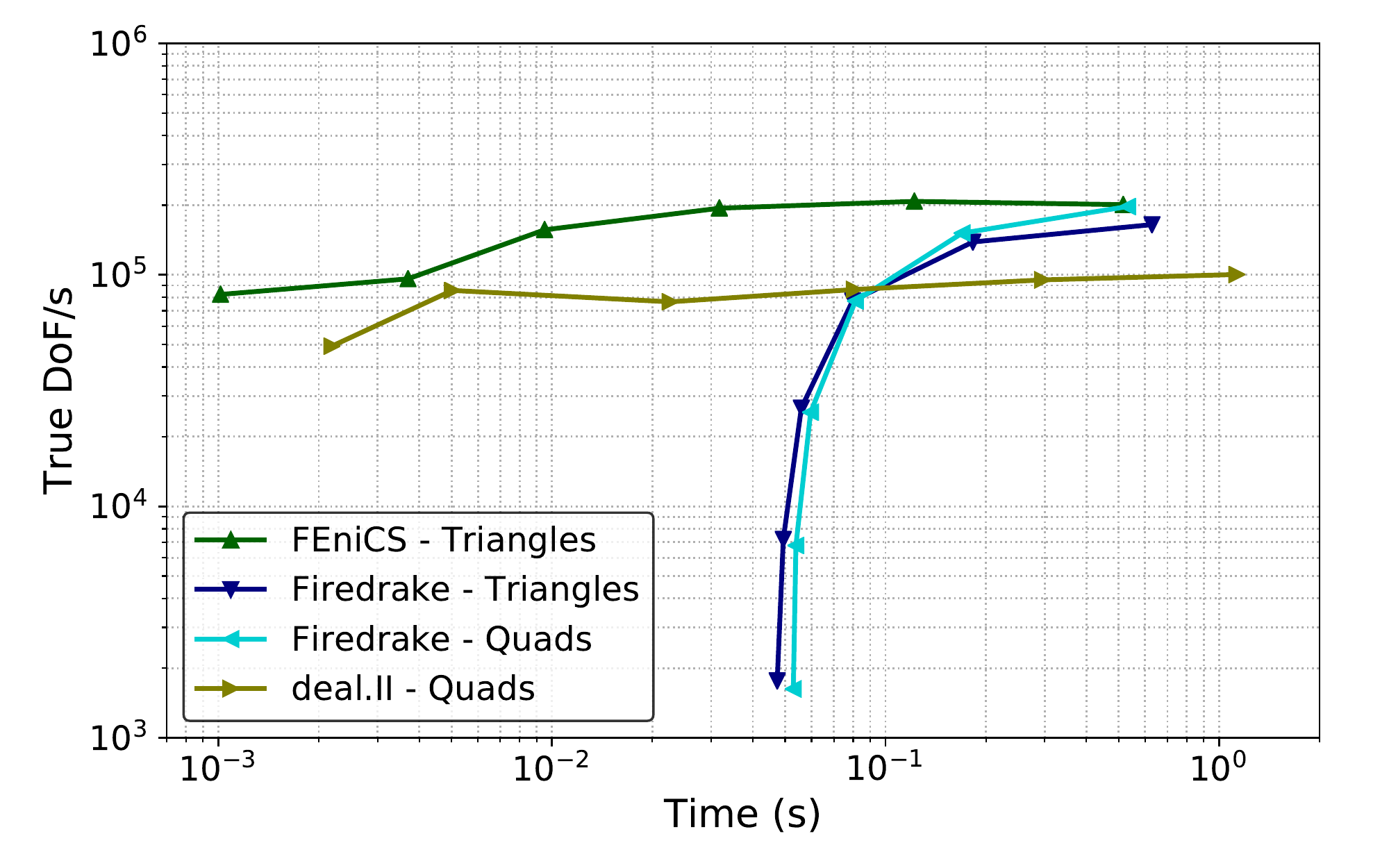}}
  \caption{Test \#1: Accuracy rates comparing various finite element software 
  packages on structured and unstructured 2D grids.\label{Fig:Test1efficacy}}
\end{figure}

First, we examine how the first-order CG discretization performs on both 
structured and unstructured grids as implemented in the three different software
packages. Consider the following analytical solution on a unit square,
%--------------------;
%  Equation: Test 1  ;
%--------------------;
\begin{align}\label{Eqn:test1solution}
  u(x, y) = \sin(2\pi x) \sin(2\pi y).
\end{align}
The 2D initial coarse grids shown in Figure~\ref{Fig:meshes} and are
refined up to 5 times. Information concerning the DoS and DoA for 
these problems can be found in Table~\ref{Tab:test1}. It should be noted that
while the Firedrake library is capable of handling both triangular and quadrilateral 
elements, the FEniCS and deal.II libraries are only capable of handling, respectively, 
triangular and quadrilateral elements.

Figure~\ref{Fig:Test1meshstatic} contain the mesh convergence and static-scaling
diagrams. From the mesh-convergence diagrams, we see that the unit slope matches 
our prediction $\frac{\alpha}{d} = 1$ for a second order method in two 
dimensions. The static-scaling diagrams indicate that that FEniCS/Dolfin and Firedrake 
have very similar performances and outperform deal.II on large problems.
However, as the problem size decreases (approaching the strong-scaling limit) the
overhead in Firedrake becomes apparent. Note that deal.II and Firedrake's 
quadrilateral meshes give slightly more accurate solutions so we now examine the two
accuracy rate metrics in Figure~\ref{Fig:Test1efficacy}. The DoE plots
indicate that the quadrilateral meshes are in fact the more accurate methods, especially
for the structured grids. The true static-scaling plots indicate that while the deal.II results may
be the most accurate, they are not the fastest in terms of processing the DoFs. 
The TAS spectrum analysis suggests that quadrilateral meshes not only offer more 
accuracy but are also faster so long as the software is implemented efficiently.

%----------------------------;
%  Subsection: Test 2  ;
%----------------------------;
\subsection{Test \#2: CG vs DG with same $h$-size}
%-------------------;
%  Figures: Test 2  ;
%-------------------;
\begin{figure}[t]
  \centering
  \subfloat[Mesh- convergence: FEniCS]{\includegraphics[width=0.4\textwidth]{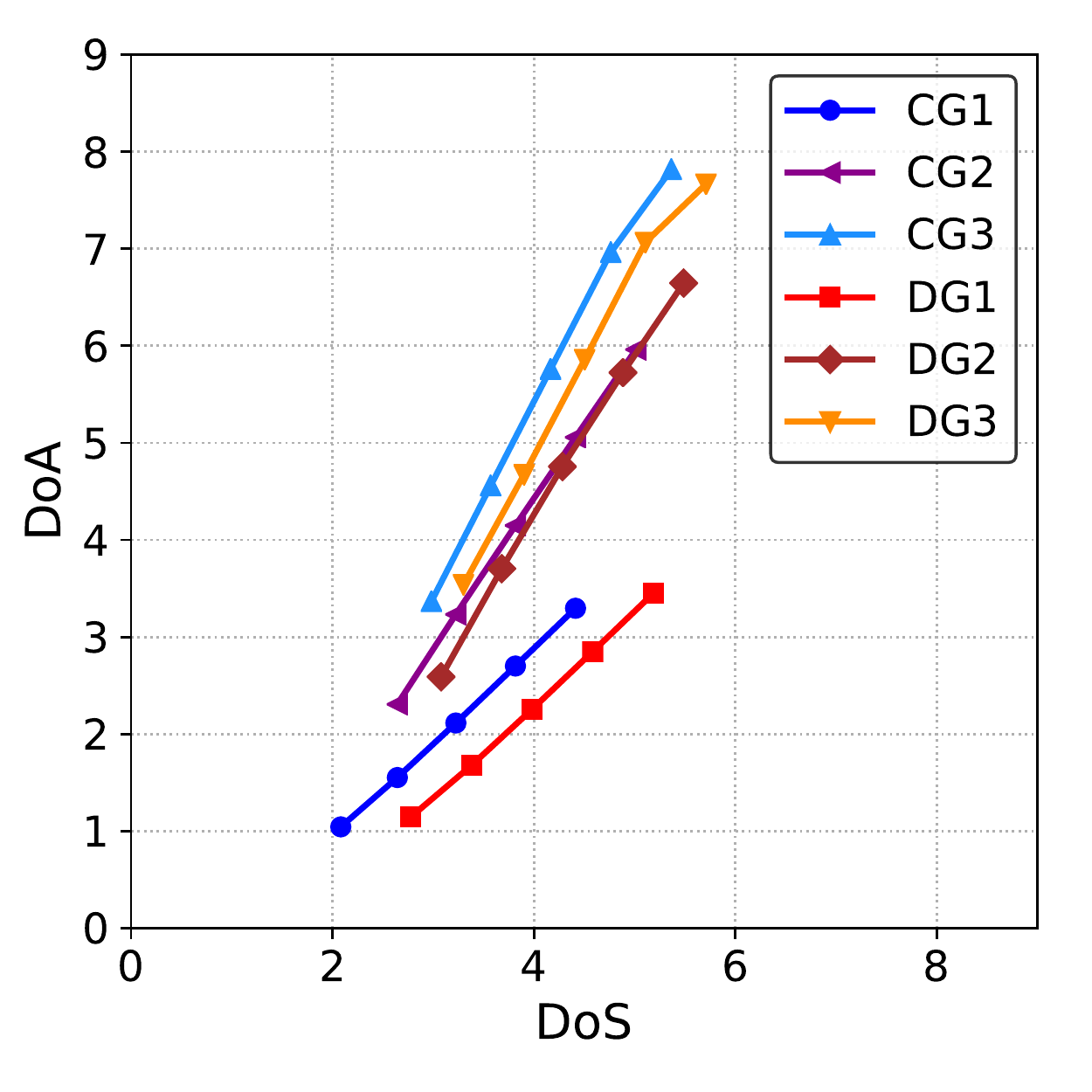}}
  \subfloat[Mesh-convergence: deal.II]{\includegraphics[width=0.4\textwidth]{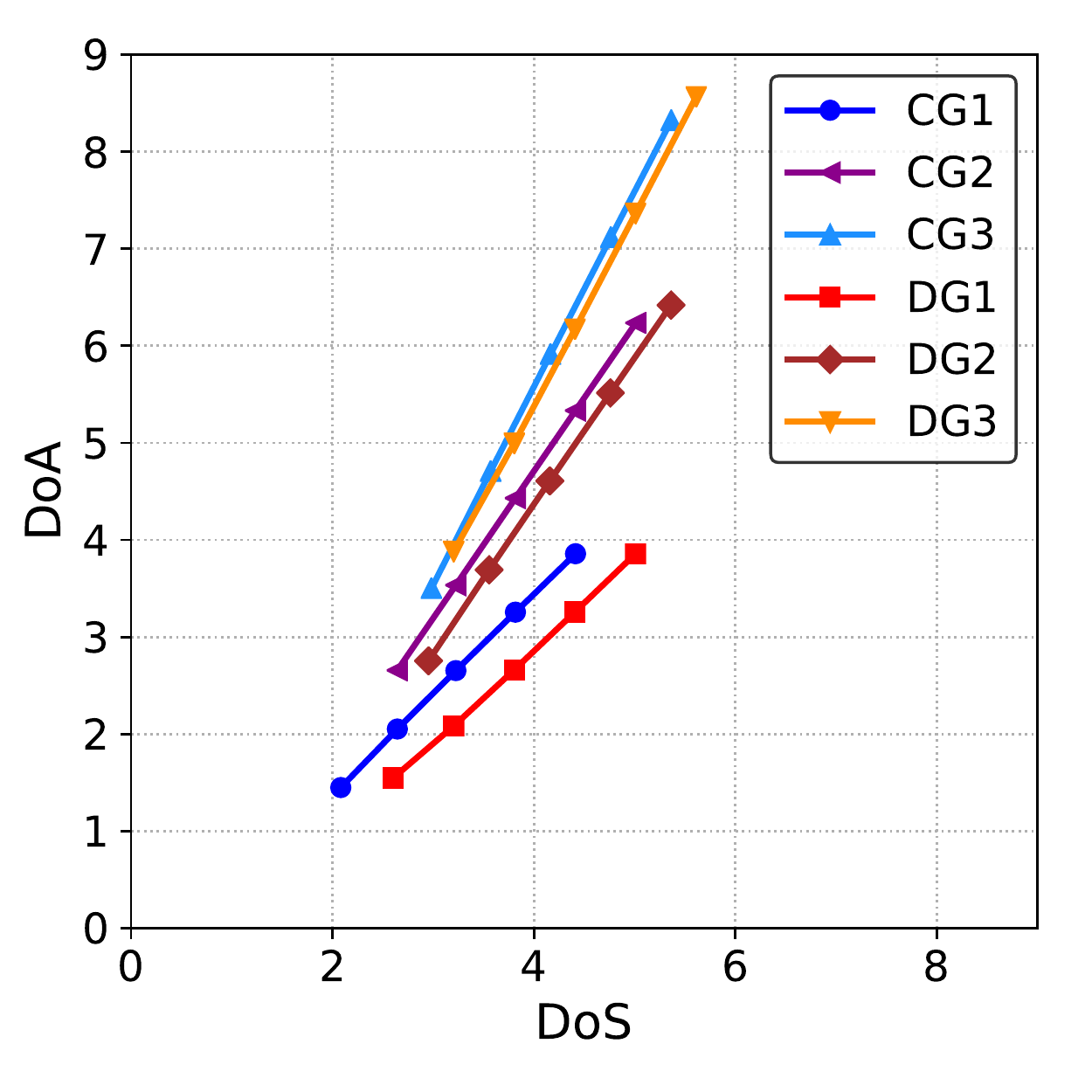}}\\
  \subfloat[Static-scaling: FEniCS]{\includegraphics[width=0.5\textwidth]{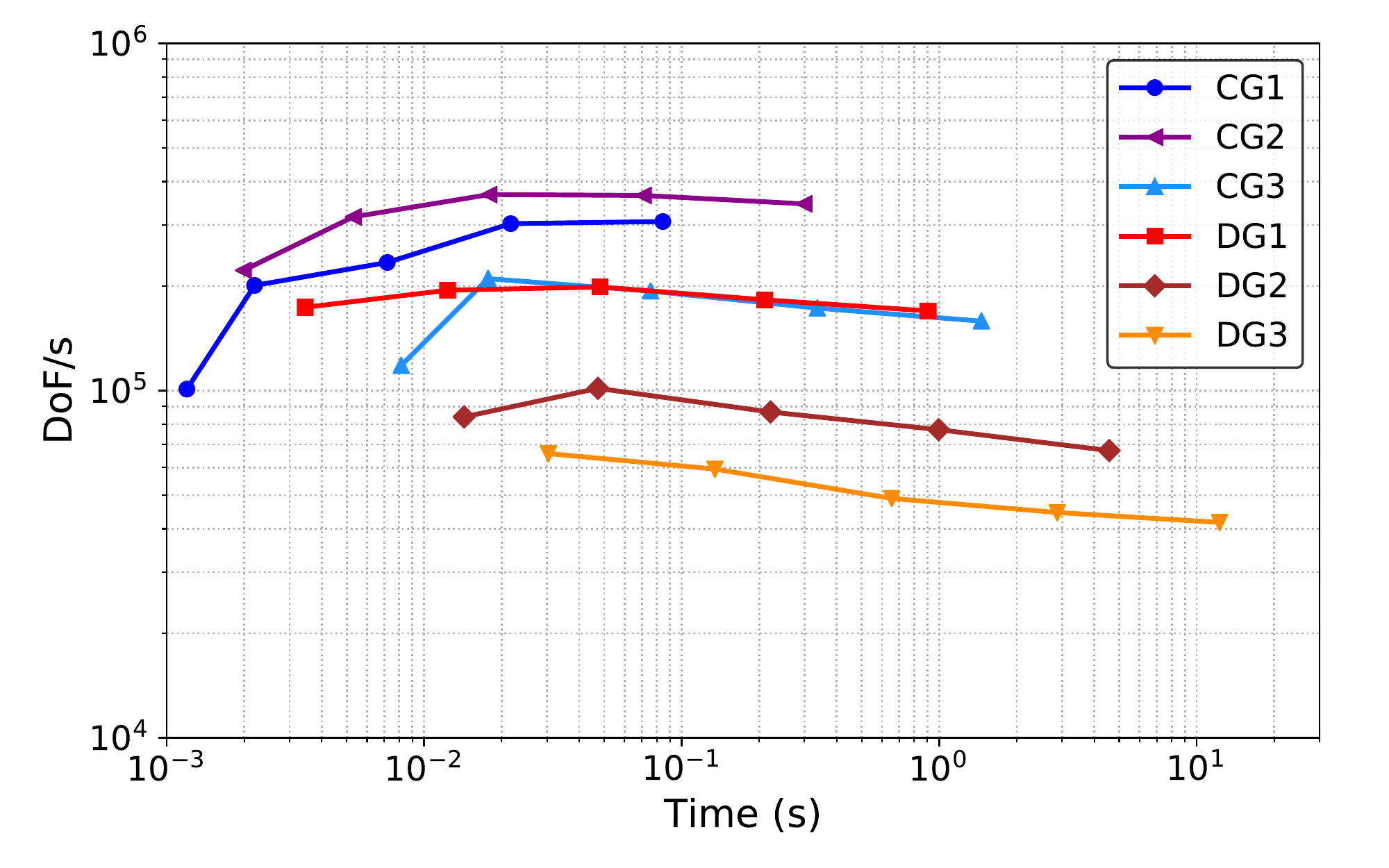}}
  \subfloat[Static-scaling: deal.II]{\includegraphics[width=0.5\textwidth]{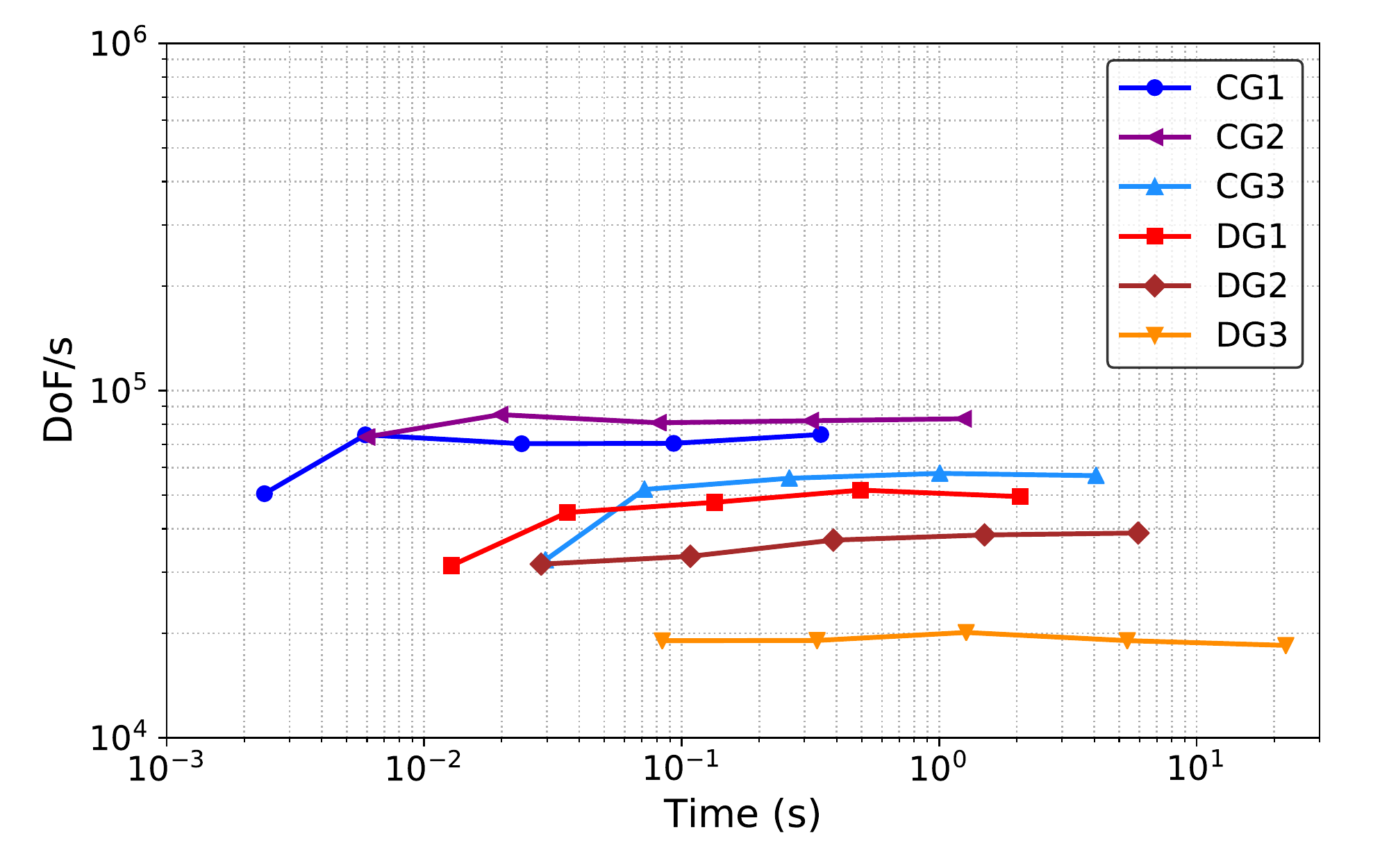}}
  \caption{Test \#2: Mesh convergence and static-scaling on structured grids comparing 2D CG and DG when the $h$-size is the same.\label{Fig:Test2meshstatic}}
\end{figure}
%--------------------;
%  Table: Test 2  ;
%--------------------;
{\tiny \begin{table}[b]
\center
\caption{Comparison of 2D structured grid (same $h$) CG and DG in FEniCS\label{Tab:test2_fenics}}
\begin{tabular}{l|ccc|ccc|ccc}
\hline
\multirow{2}{*}{$h$-size} & \multicolumn{3}{c|}{CG1} & \multicolumn{3}{c|}{CG2} &
  \multicolumn{3}{c}{CG3} \\
& DoA & DoS & DoA/DoS & DoA & DoS & DoA/DoS &DoA & DoS & DoA/DoS  \\
\hline
1/10 & 1.04 & 2.08 & 0.50 & 2.31 & 2.64 & 0.88 & 3.37 & 2.98 & 1.13 \\
1/20 & 1.55 & 2.64 & 0.59 & 3.23 & 3.23 & 1.00 & 4.56 & 3.57 & 1.28 \\
1/40 & 2.11 & 3.23 & 0.65 & 4.15 & 3.82 & 1.09 & 5.76 & 4.17 & 1.38 \\
1/80 & 2.70 & 3.82 & 0.71 & 5.06 & 4.41 & 1.15 & 6.96 & 4.76 & 1.46 \\
1/160 & 3.30 & 4.41 & 0.75 & 5.96 & 5.01 & 1.19 & 7.82 & 5.36 & 1.45 \\
\hline
\hline
\multirow{2}{*}{$h$-size} & \multicolumn{3}{c|}{DG1} & \multicolumn{3}{c|}{DG2} &
  \multicolumn{3}{c}{DG3} \\
& DoA & DoS & DoA/DoS & DoA & DoS & DoA/DoS &DoA & DoS & DoA/DoS  \\
\hline
1/10 & 1.15 & 2.78 & 0.41 & 2.59 & 3.08 & 0.84 & 3.54 & 3.30 & 1.07 \\
1/20 & 1.68 & 3.38 & 0.50 & 3.70 & 3.68 & 1.01 & 4.67 & 3.90 & 1.20 \\
1/40 & 2.26 & 3.98 & 0.57 & 4.76 & 4.28 & 1.11 & 5.86 & 4.51 & 1.30 \\
1/80 & 2.85 & 4.58 & 0.62 & 5.72 & 4.89 & 1.17 & 7.07 & 5.11 & 1.38 \\
1/160 & 3.45 & 5.19 & 0.66 & 6.65 & 5.49 & 1.21 & 7.67 & 5.71 & 1.34 \\
\hline
\end{tabular}
\end{table}
\begin{table}[b]
\center
\caption{Comparison of 2D structured grid (same $h$) CG and DG in deal.II\label{Tab:test2_dealii}}
\begin{tabular}{l|ccc|ccc|ccc}
\hline
\multirow{2}{*}{$h$-size} & \multicolumn{3}{c|}{CG1} & \multicolumn{3}{c|}{CG2} &
  \multicolumn{3}{c}{CG3} \\
& DoA & DoS & DoA/DoS & DoA & DoS & DoA/DoS &DoA & DoS & DoA/DoS  \\
\hline
1/10 & 1.45 & 2.08 & 0.70 & 2.65 & 2.64 & 1.00 & 3.50 & 2.98 & 1.17 \\
1/20 & 2.05 & 2.64 & 0.88 & 3.53 & 3.23 & 1.10 & 4.71 & 3.57 & 1.32 \\
1/40 & 2.65 & 3.23 & 0.82 & 4.43 & 3.82 & 1.16 & 5.91 & 4.17 & 1.42 \\
1/80 & 3.26 & 3.82 & 0.85 & 5.33 & 4.41 & 1.21 & 7.12 & 4.76 & 1.49 \\
1/160& 3.86 & 4.41 & 0.87 & 6.24 & 5.01 & 1.24 & 8.32 & 5.36 & 1.55 \\
\hline
\hline
\multirow{2}{*}{$h$-size} & \multicolumn{3}{c|}{DG1} & \multicolumn{3}{c|}{DG2} &
  \multicolumn{3}{c}{DG3} \\
     & DoA  & DoS &DoA/DoS& DoA  & DoS  &DoA/DoS & DoA  & DoS  & DoA/DoS  \\
\hline
1/10 & 1.55 & 2.60 & 0.59 & 2.75 & 2.95 & 0.93   & 3.88 & 3.20 & 1.21 \\
1/20 & 2.08 & 3.20 & 0.65 & 3.69 & 3.56 & 1.04   & 5.00 & 3.81 & 1.31 \\
1/40 & 2.66 & 3.81 & 0.70 & 4.61 & 4.16 & 1.11   & 6.17 & 4.41 & 1.40 \\
1/80 & 3.26 & 4.41 & 0.74 & 5.52 & 4.76 & 1.16   & 7.37 & 5.01 & 1.47 \\
1/160& 3.86 & 5.01 & 0.77 & 6.42 & 5.36 & 1.20   & 8.57 & 5.61 & 1.53 \\
\hline
\end{tabular}
\end{table}
}
\begin{figure}[t]
  \centering
  \subfloat[DoE: FEniCS]{\includegraphics[width=0.5\textwidth]{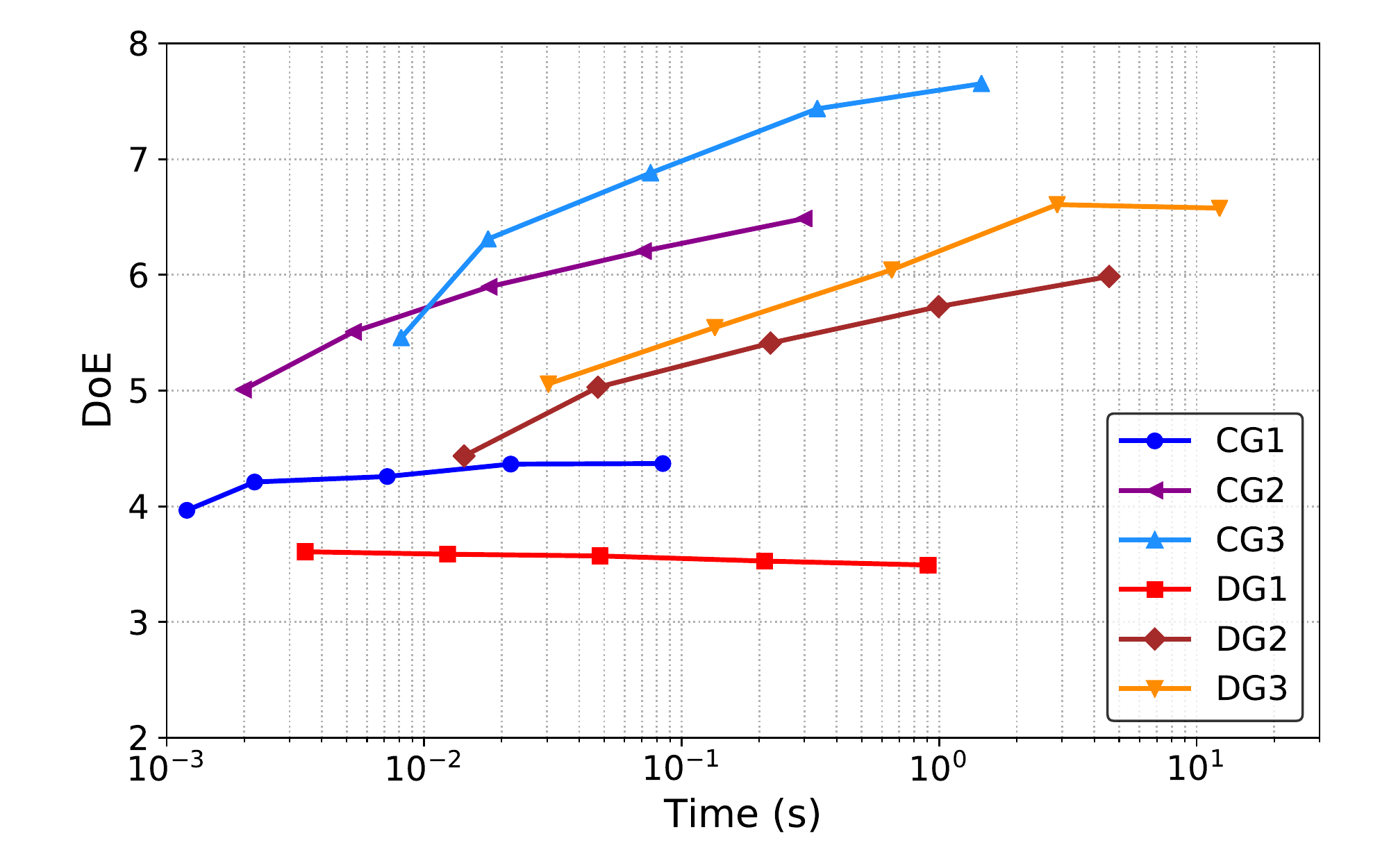}}
  \subfloat[DoE: deal.II]{\includegraphics[width=0.5\textwidth]{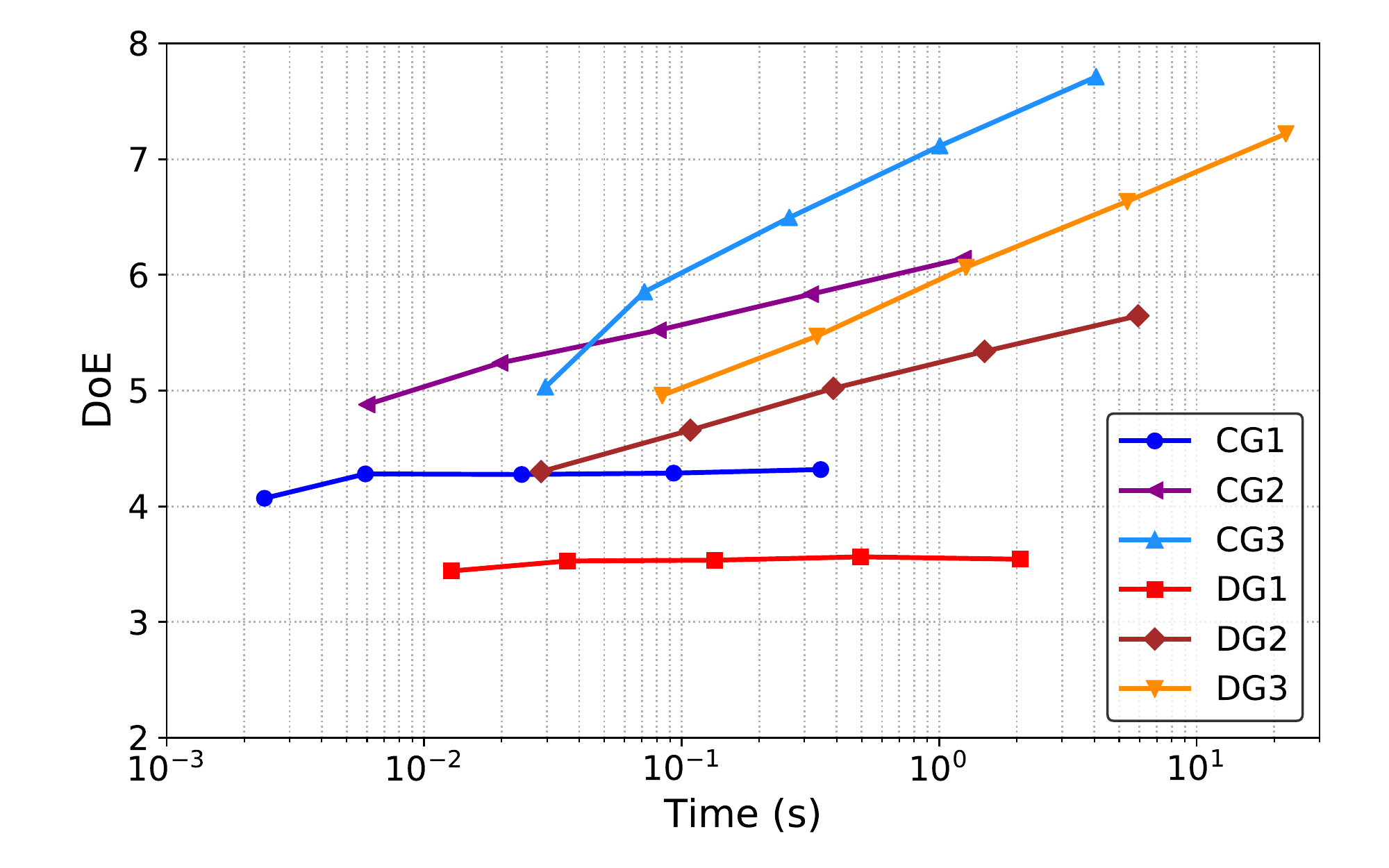}}\\
  \subfloat[True static-scaling: FEniCS]{\includegraphics[width=0.5\textwidth]{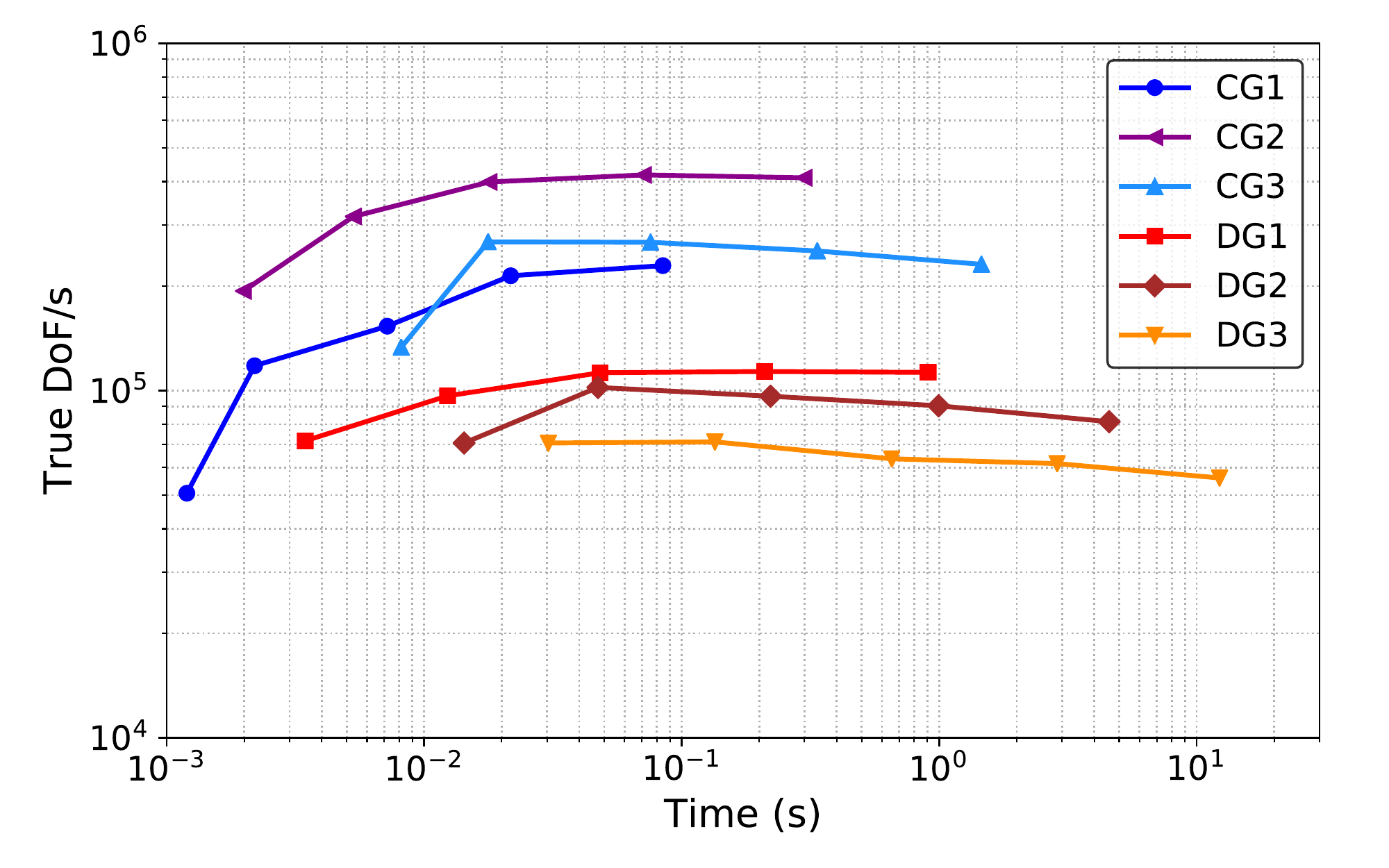}}
  \subfloat[True static-scaling: deal.II]{\includegraphics[width=0.5\textwidth]{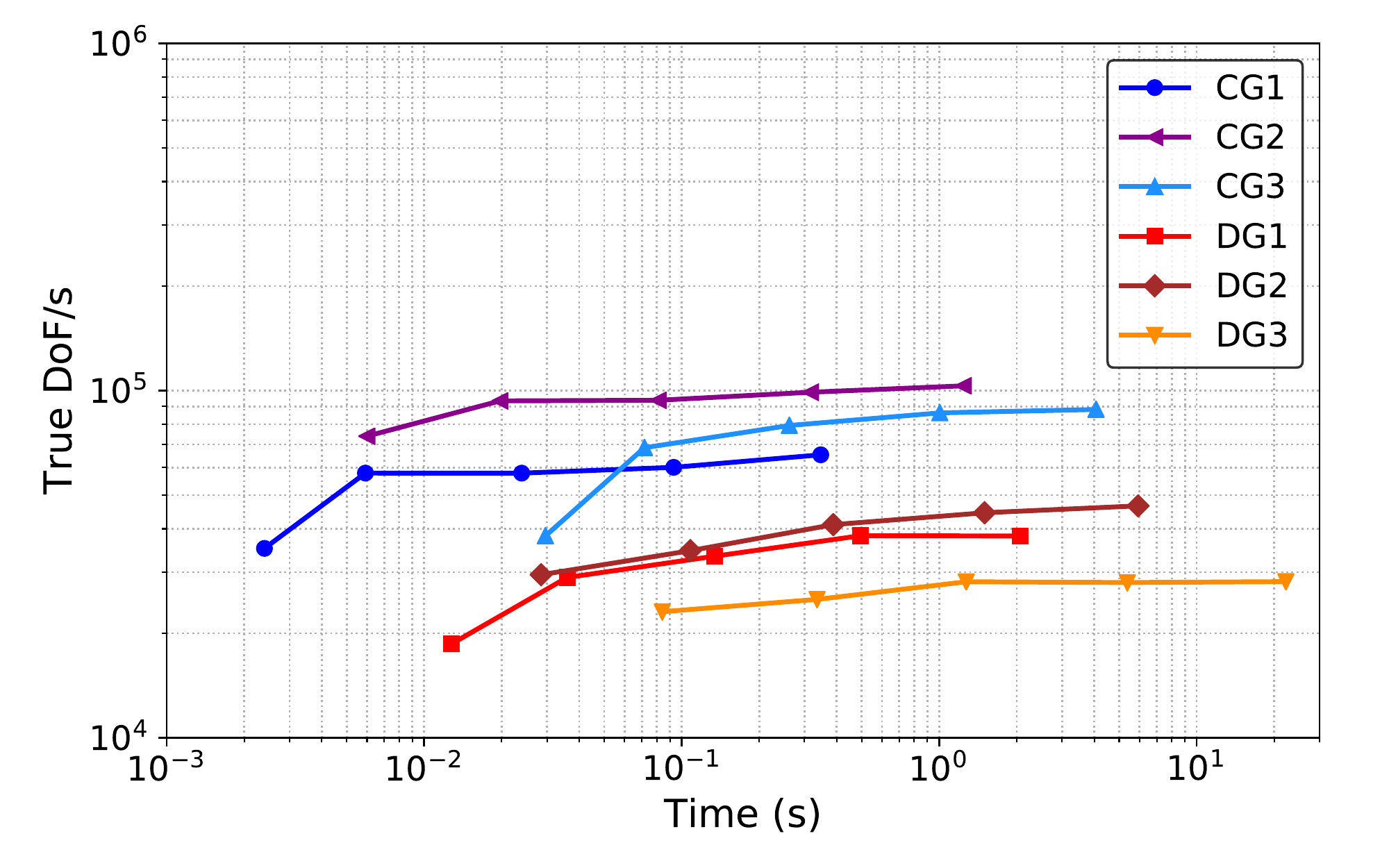}}
  \caption{Test \#2: Accuracy rates on structured grids comparing 2D CG and DG when the $h$-size is the same.\label{Fig:Test2efficacy}}
\end{figure}

Next, we examine the behavior of different finite element discretizations for a given 
refinement, or $h$ size. Three levels of $p$-refinement are considered on a unit square 
domain with the following analytical solution,
\begin{align}\label{Eqn:test2solution}
  u(x,y) = \sin(2\pi x^{2}) \sin(2\pi y^{2}).
\end{align}
We consider only the CG and SIP DG discretizations, with the value of the penalty chosen based 
on the formula derived in~\cite{Shahbazi2005}. Both FEniCS/Dolfin and deal.II are used for this case, starting with the 
structured meshes on the left of Figure~\ref{Fig:meshes} as the initial meshes and using four levels 
of refinement. The DoS and DoA for the FEniCS/Dolfin and deal.II discretizations can be found in 
Tables~\ref{Tab:test2_fenics} and \ref{Tab:test2_dealii}, respectively. From 
the mesh-convergence diagrams in Figure~\ref{Fig:Test2meshstatic}, we see that second order 
methods have unit slope, the third order methods have slope $\frac{3}{2}$, and fourth order methods 
have slope 2, as predicted. It is interesting to note that FEniCS's fourth order methods experience a dropoff 
in convergence when the DoA goes past 7.0, which arises due to the relative convergence criterion
of 10$^{-7}$.

In the static-scaling diagrams in Figure~\ref{Fig:Test2meshstatic}, we see that the high order methods 
show greater fall off as problem size increases. Since the number of solver iterates remains 
roughly constant, this is due to AMG solver complexity rising at a slightly nonlinear rate. 
Moreover, the CG methods perform at a strictly higher rate than their DG counterparts, and 
within the CG and DG classes, lower order methods are operating 
faster than high order, with the exception of CG2. When
we introduce the notion of accuracy into Figure~\ref{Fig:Test2efficacy}, however, 
this traditional analysis is upended. Now, within CG or DG, each order
produces accuracy faster than the order below as seen from the scaled error diagrams. 
From the true static-scaling diagrams, it can be seen that the CG methods are slightly 
more efficient. For example, DG3 may be one of the most accurate methods for smaller $h$-sizes
but is clearly the slowest at processing its DoFs if all DoFs are given equal treatment. 

%----------------------------;
%  Subsection: Test 3  ;
%----------------------------;
\subsection{Test \#3: CG vs DG with same DoF count}
%-------------------;
%  Figures: Test 3  ;
%-------------------;
\begin{figure}[t]
  \centering
  \subfloat[Mesh-convergence: Tetrahedrons]{\includegraphics[width=0.4\textwidth]{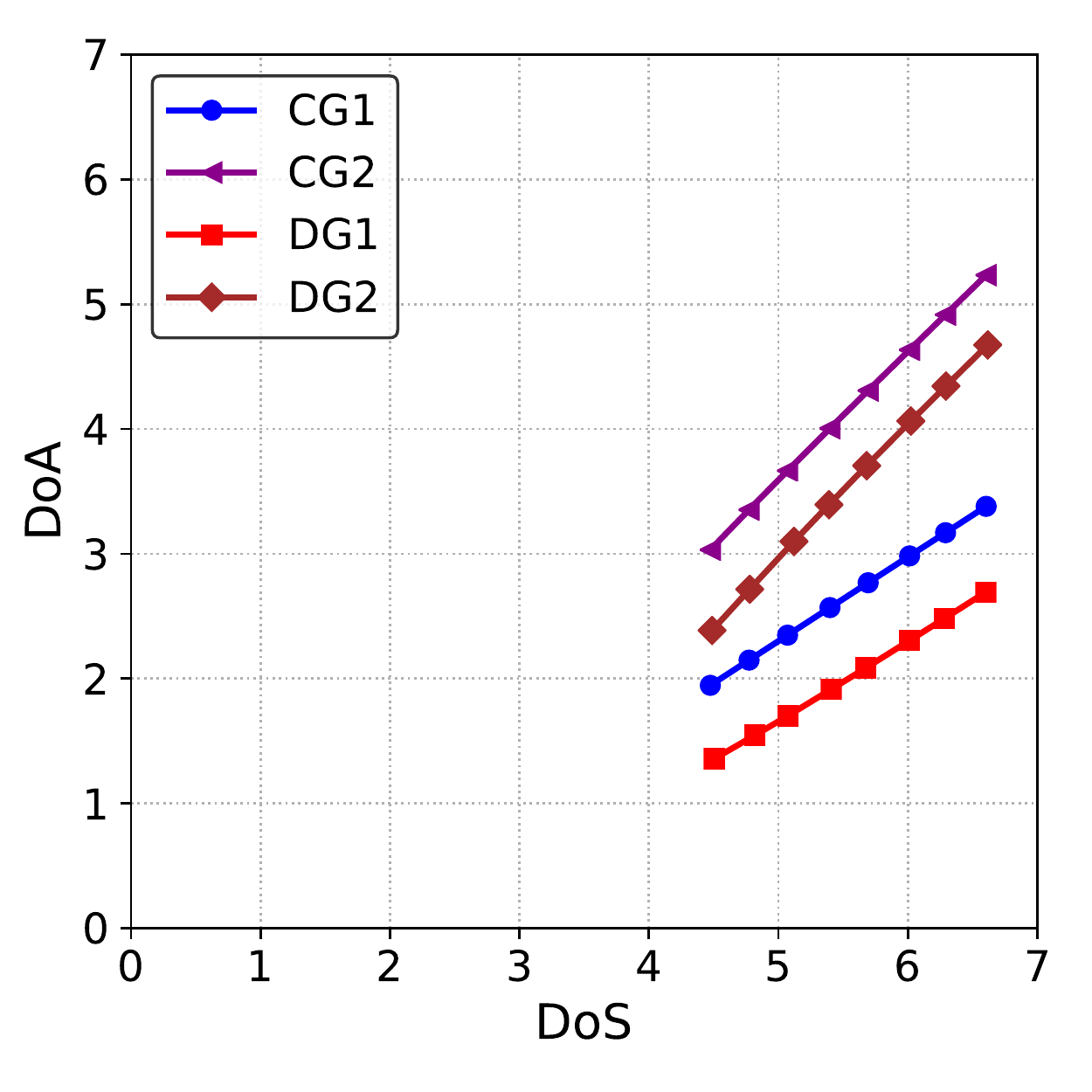}}
  \subfloat[Mesh-convergence: Hexahedrons]{\includegraphics[width=0.4\textwidth]{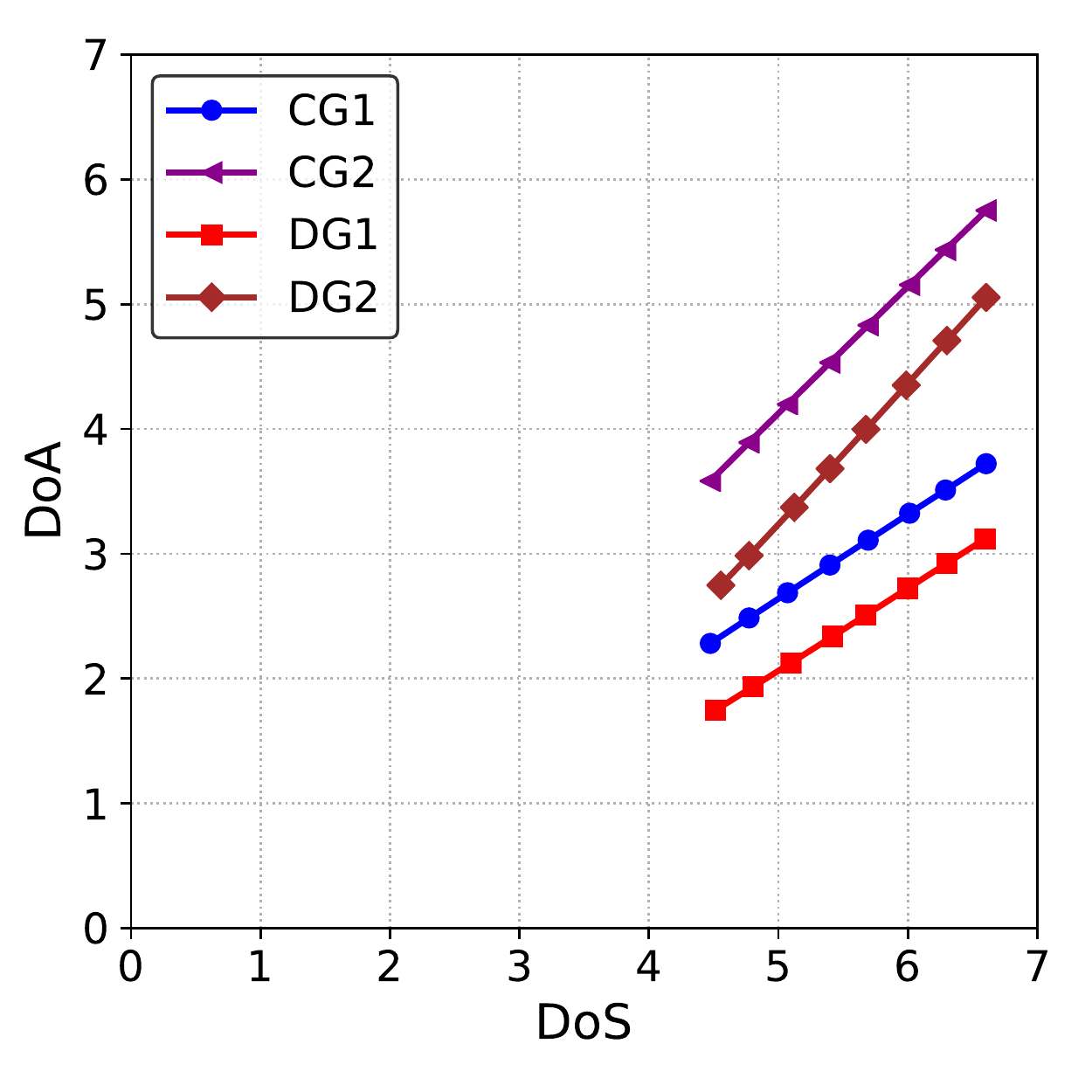}}\\
  \subfloat[Static-scaling: Tetrahedrons]{\includegraphics[width=0.5\textwidth]{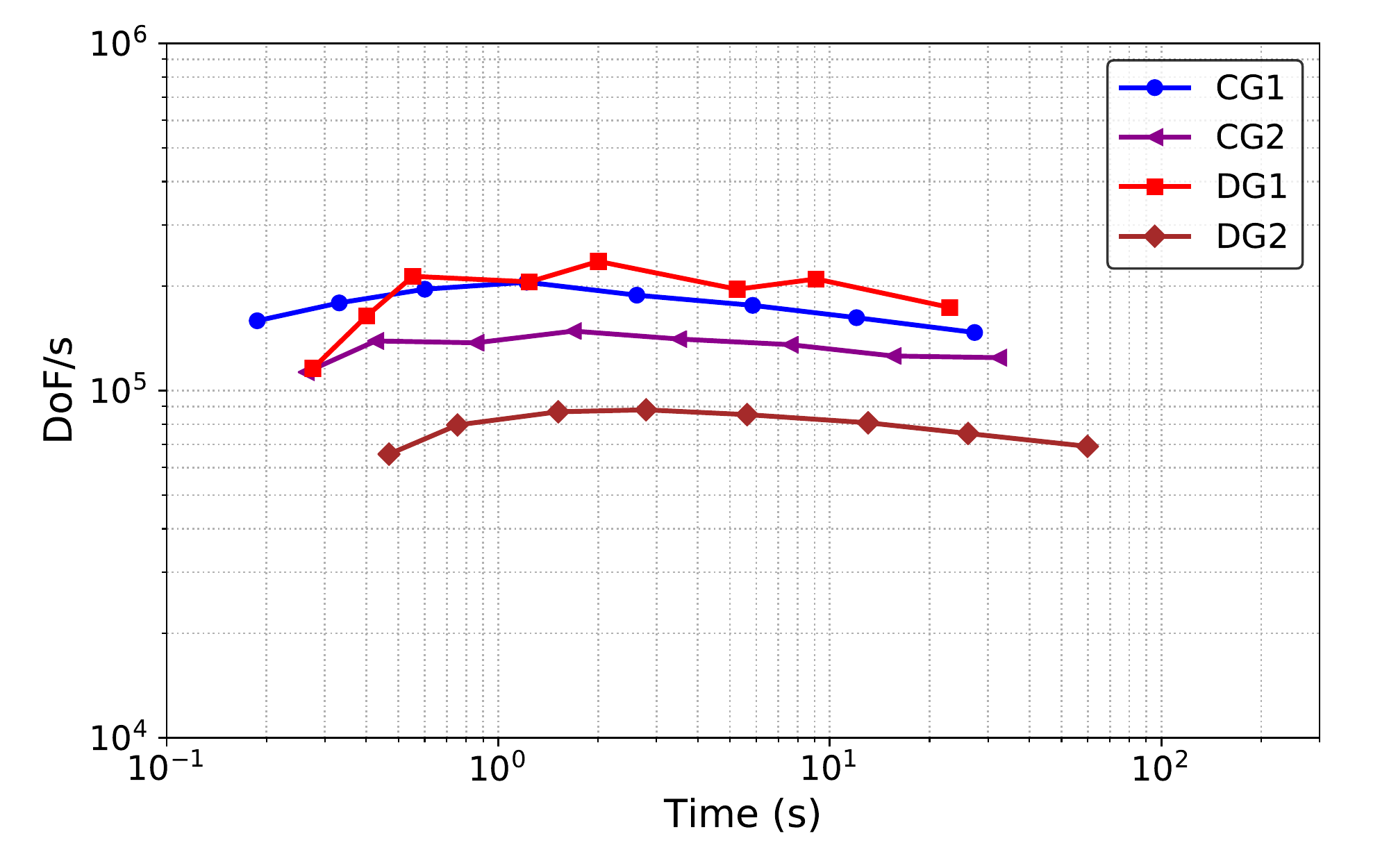}}
  \subfloat[Static-scaling: Hexahedrons]{\includegraphics[width=0.5\textwidth]{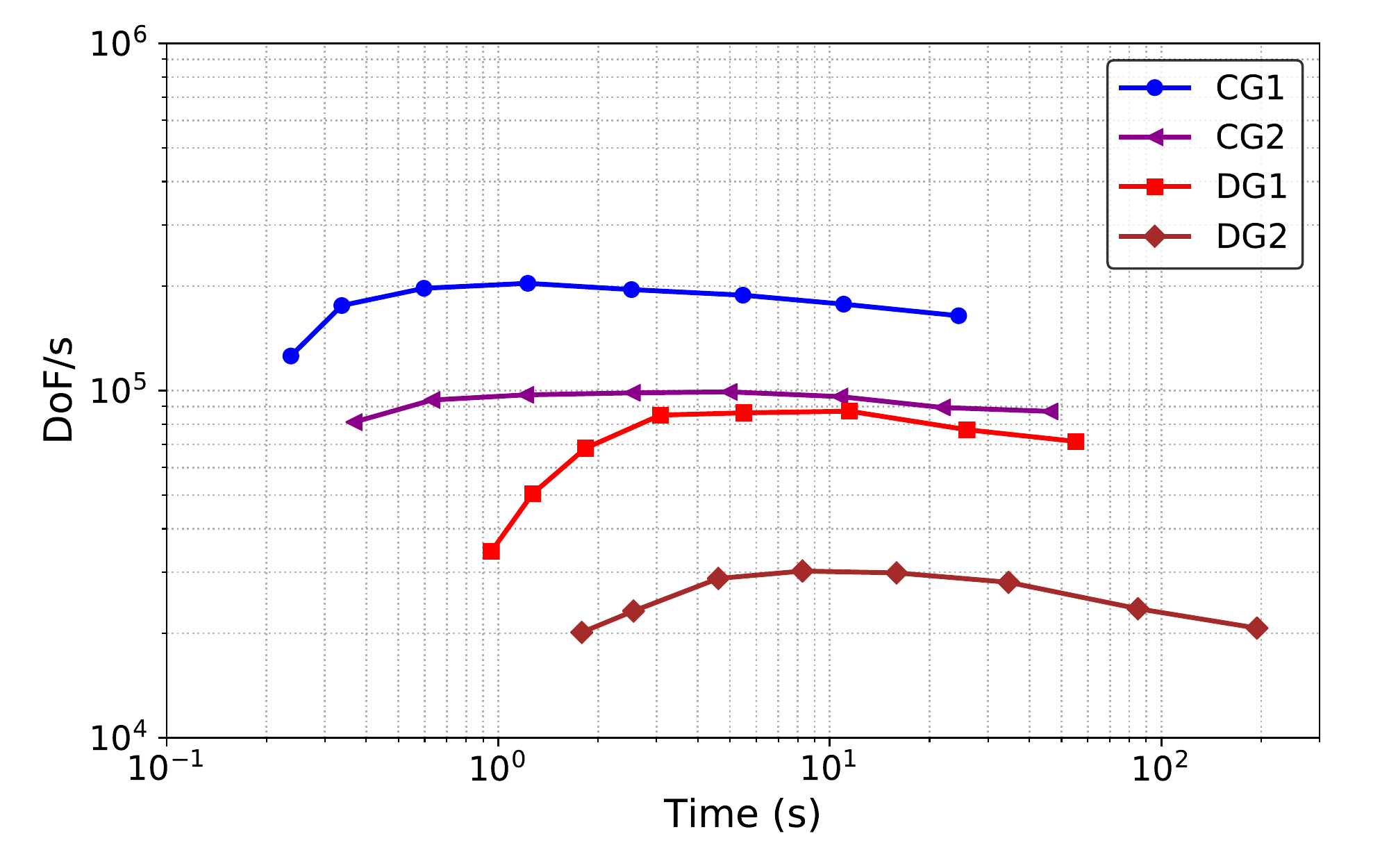}}
  \caption{Test \#3: Mesh convergence and static-scaling on structured grids
  comparing CG and DG when the DoF is the same.\label{Fig:Test3meshstatic}}
\end{figure}
\begin{figure}[t]
  \centering
  \subfloat[DoE: Tetrahedrons]{\includegraphics[width=0.5\textwidth]{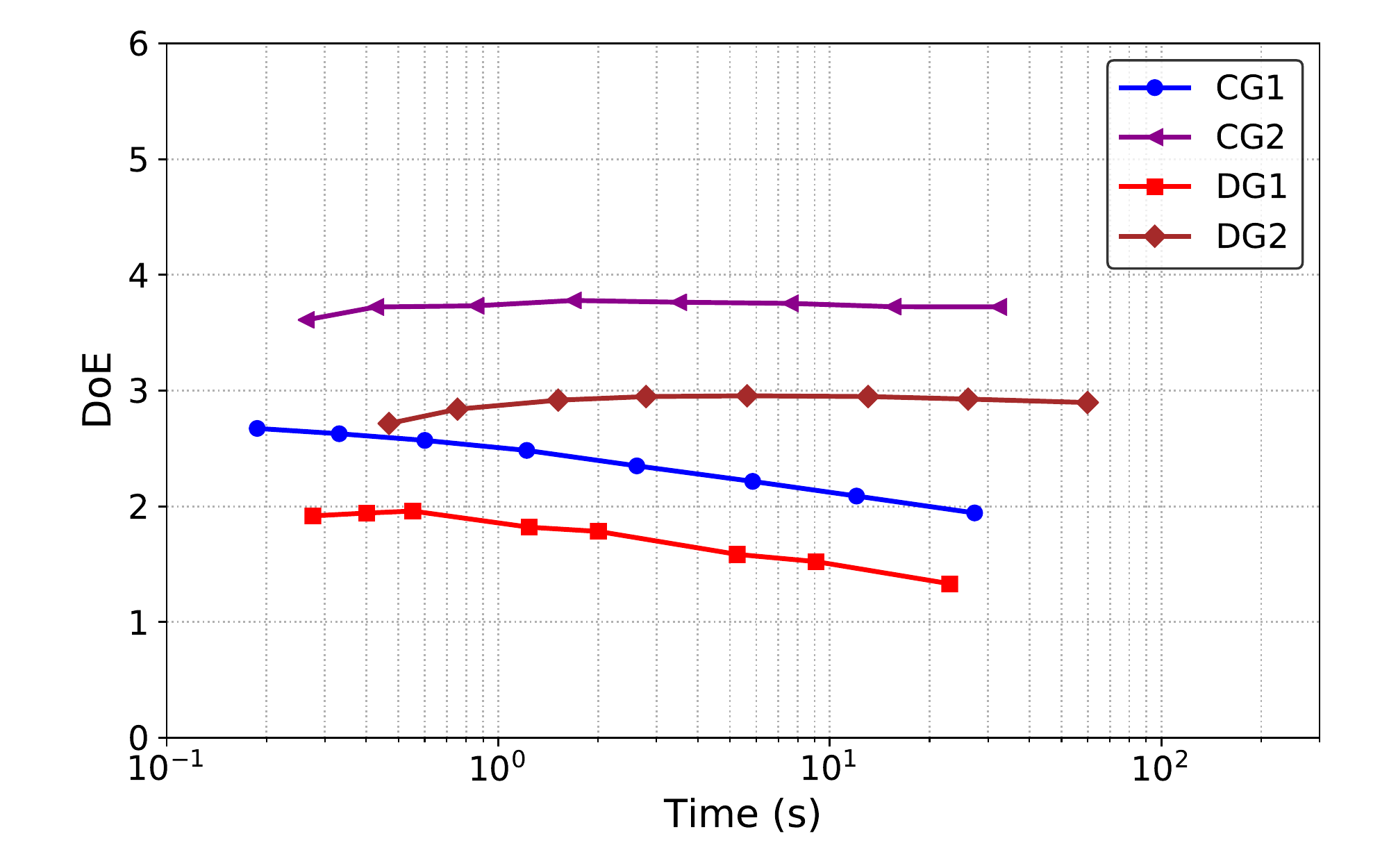}}
  \subfloat[DoE: Hexahedrons]{\includegraphics[width=0.5\textwidth]{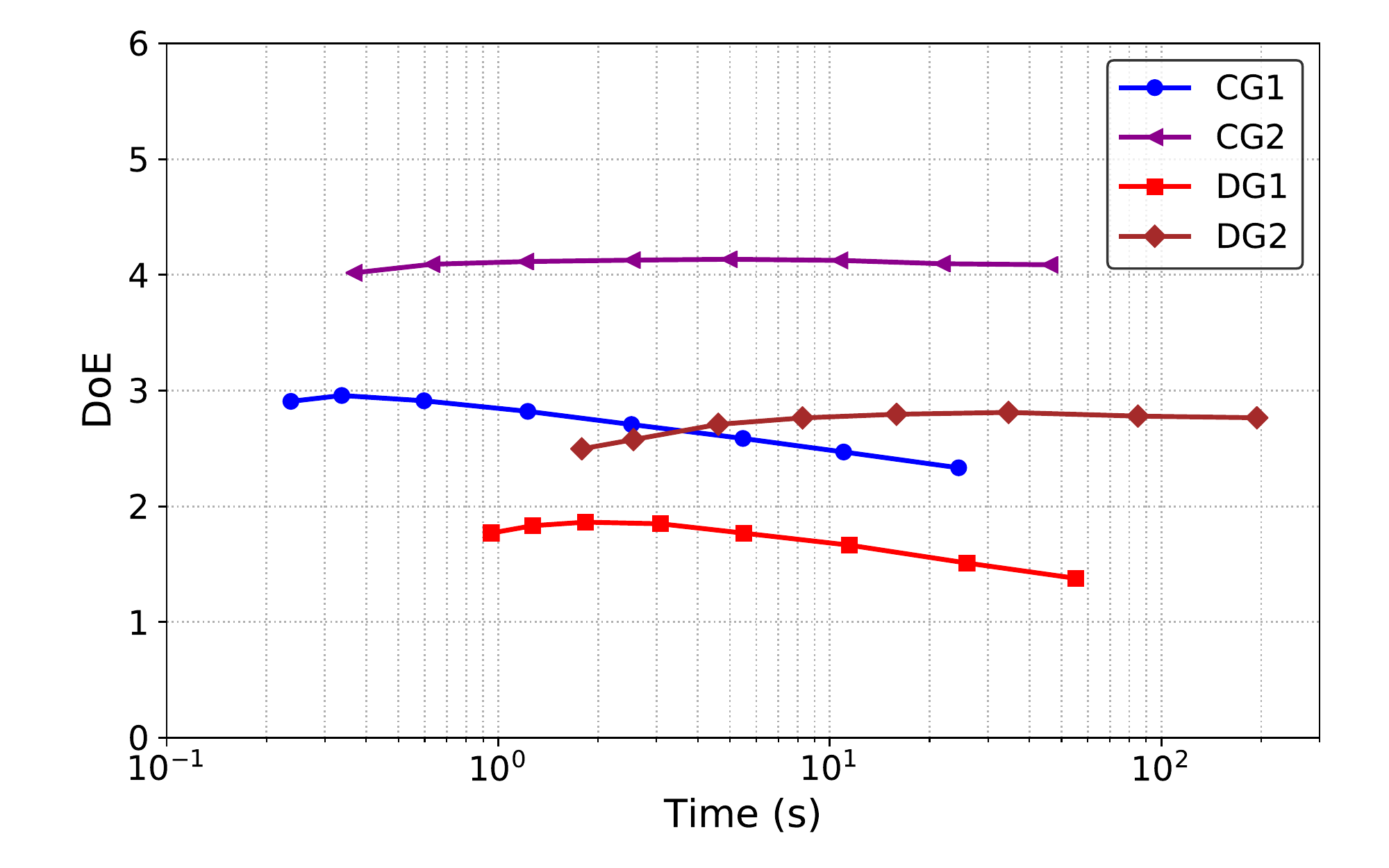}}\\
  \subfloat[True static-scaling: Tetrahedrons]{\includegraphics[width=0.5\textwidth]{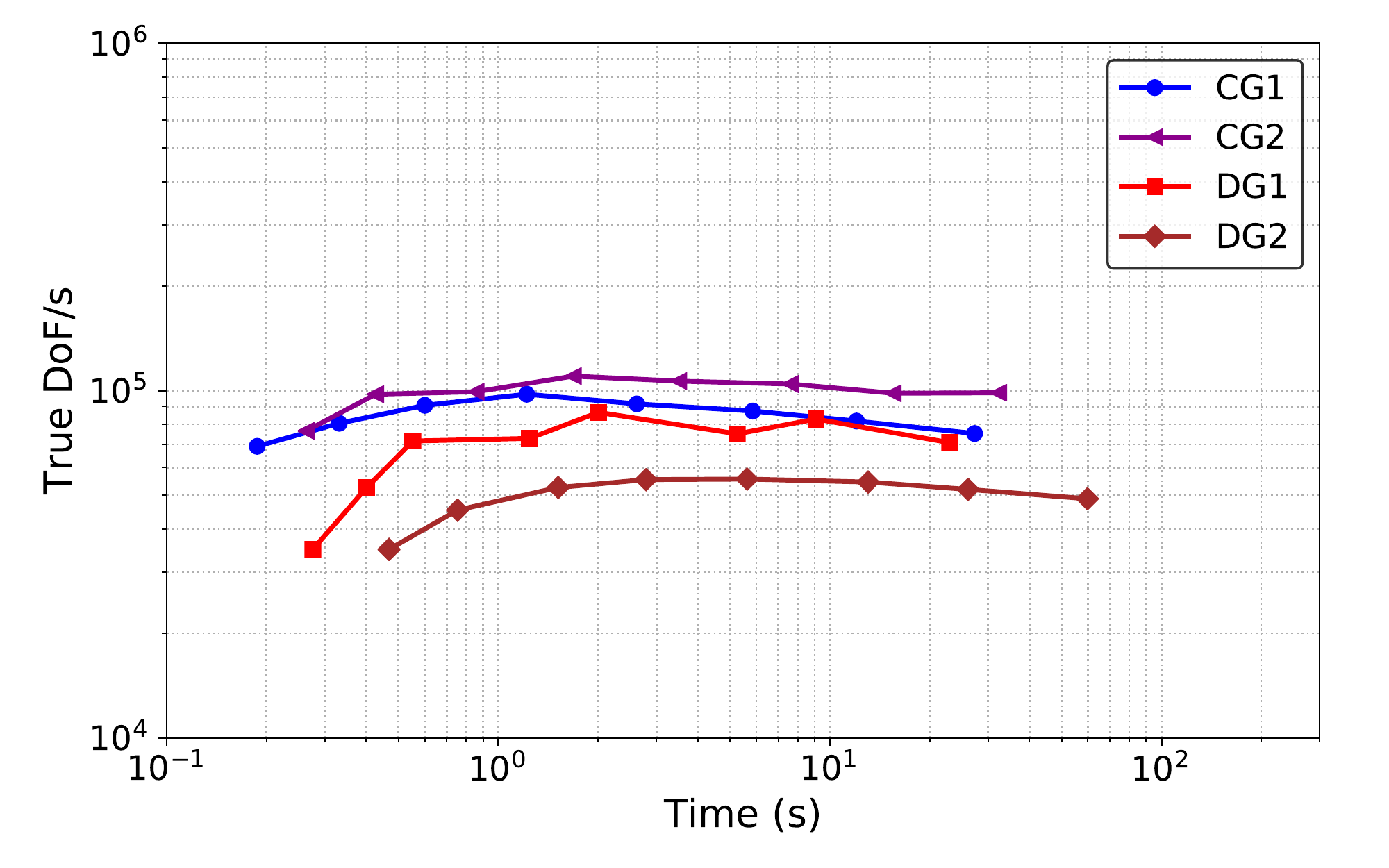}}
  \subfloat[True static-scaling: Hexahedrons]{\includegraphics[width=0.5\textwidth]{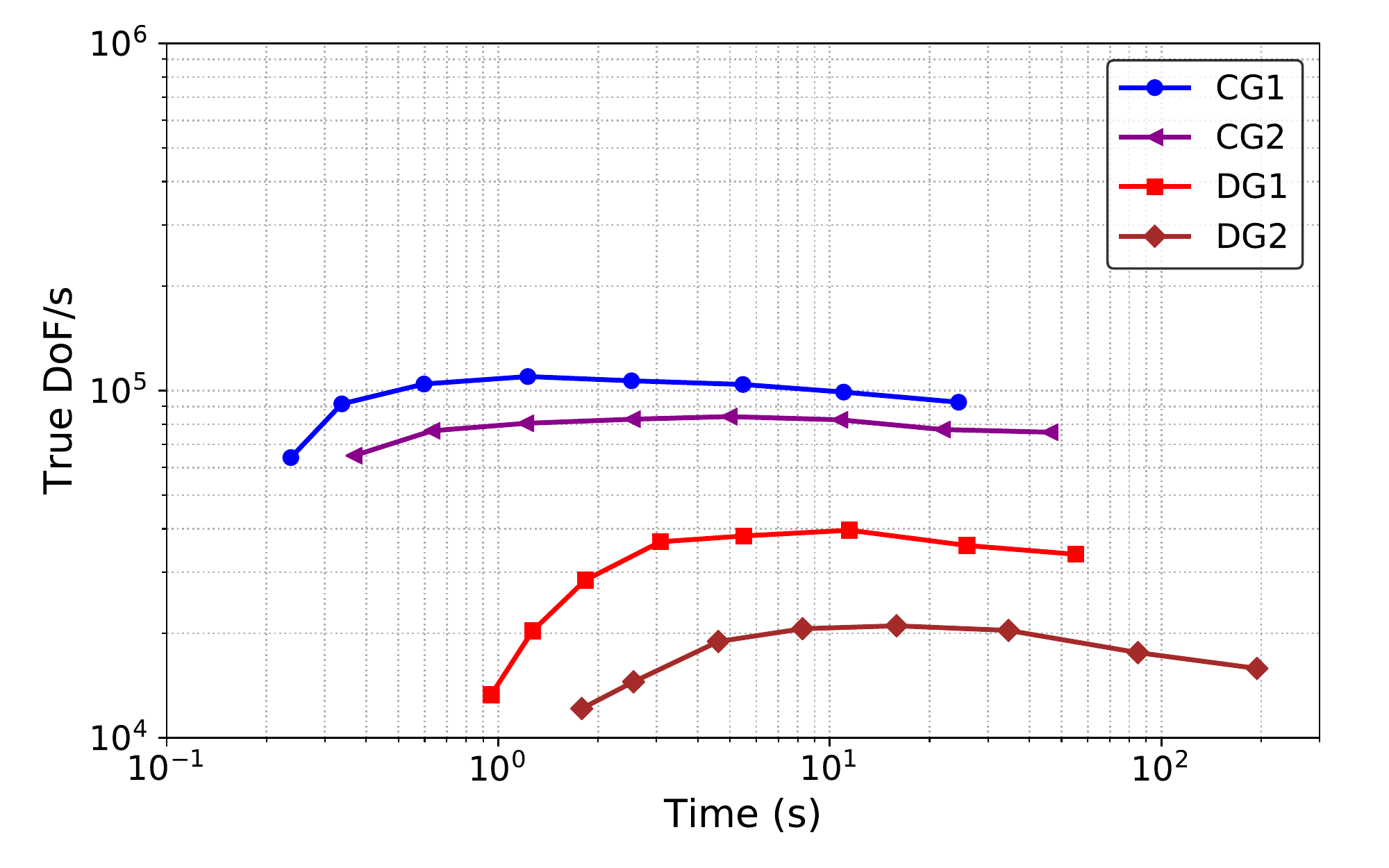}}
  \caption{Test \#3: Accuracy rates on structured grids comparing 
  3D CG and DG when the DoF the same.\label{Fig:Test3efficacy}}
\end{figure}

It can be seen from the previous test that each discretization has a different DoF count for a given mesh,
making it somewhat difficult to draw any comparisons. This can be especially true for 3D problems where the
problem size proliferates even more for every step of mesh refinement. In this third test, each 
finite element discretization will have roughly the same DoF count by adjusting the $h$-size. Using 
Firedrake, the CG and DG discretizations with up to 2 different levels of $p$-refinement 
are considered for the following analytical solution on a unit cube,
\begin{align}\label{Eqn:test3solution}
  u(x, y, z) = \sin(2\pi x) \sin(2\pi y) \sin(2\pi z).
\end{align}
Both tetrahedral and hexahedral elements from Figure~\ref{Fig:meshes} are used and 
Tables~\ref{Tab:test3_tets} and \ref{Tab:test3_hexes} depict the $h$-sizes needed in order for all the discretizations to 
have roughly the same DoS. From Figure~\ref{Fig:Test3meshstatic}, we see that second 
order methods have slope $\frac{2}{3}$, since now $d = 3$, and the third order methods 
have unit slope, as predicted. 
{\tiny \begin{table}[t]
\center
\caption{Comparison of 3D structured grid with tetrahedron elements (same DoF) CG and DG in Firedrake\label{Tab:test3_tets}}
\begin{tabular}{cccc|cccc}
\hline
\multicolumn{4}{c|}{CG1} & \multicolumn{4}{c}{CG2} \\
$h$-size & DoA & DoS & DoA/DoS & $h$-size & DoA & DoS & DoA/DoS  \\
\hline
1/30 & 1.96 & 4.47 & 0.43 & 1/15 & 3.03 & 4.47 & 0.68 \\
1/38 & 2.15 & 4.77 & 0.45 & 1/19 & 3.35 & 4.77 & 0.70  \\
1/48 & 2.35 & 5.07 & 0.46 & 1/24 & 3.67 & 5.07 & 0.72  \\
1/62 & 2.57 & 5.40 & 0.48 & 1/31 & 4.01 & 5.40 & 0.74  \\
1/78 & 2.77 & 5.69 & 0.49 & 1/39 & 4.31 & 5.69 & 0.76  \\
1/100 & 2.98 & 6.01 & 0.50 & 1/50 & 4.63 & 6.01 & 0.77  \\
1/124 & 3.17 & 6.29 & 0.50 & 1/62 & 4.92 & 6.29 & 0.78  \\
1/158 & 3.38 & 6.60 & 0.51 & 1/79 & 5.23 & 6.60 & 0.79  \\
\hline
\hline
\multicolumn{4}{c|}{DG1} & \multicolumn{4}{c}{DG2} \\
$h$-size & DoA & DoS & DoA/DoS & $h$-size & DoA & DoS & DoA/DoS  \\
\hline
1/11 & 1.36 & 4.50 & 0.30 & 1/8 & 2.39 & 4.49 & 0.53  \\
1/14 & 1.55 & 4.82 & 0.32 & 1/10 & 2.72 & 4.78 & 0.57  \\
1/17 & 1.70 & 5.07 & 0.34 & 1/13 & 3.10 & 5.12 & 0.61  \\
1/22 & 1.91 & 5.41 & 0.35 & 1/16 & 3.39 & 5.39 & 0.63  \\
1/27 & 2.09 & 5.67 & 0.37 & 1/20 & 3.71 & 5.68 & 0.65 \\
1/35 & 2.31 & 6.01 & 0.38 & 1/26 & 4.06 & 6.02 & 0.67 \\
1/43 & 2.48 & 6.28 & 0.40 & 1/32 & 4.34 & 6.29 & 0.69 \\
1/55 & 2.69 & 6.60 & 0.41 & 1/41 & 4.67 & 6.62 & 0.71 \\
\hline
\end{tabular}
\end{table}
\begin{table}[t]
\center
\caption{Comparison of 3D structured grid with hexahedron elements (same DoF) CG and DG in Firedrake\label{Tab:test3_hexes}}
\begin{tabular}{cccc|cccc}
\hline
\multicolumn{4}{c|}{CG1} & \multicolumn{4}{c}{CG2} \\
$h$-size & DoA & DoS & DoA/DoS & $h$-size & DoA & DoS & DoA/DoS  \\
\hline
1/30 & 2.28 & 4.47 & 0.51 & 1/15 & 3.58 & 4.47 & 0.80 \\
1/38 & 2.49 & 4.77 & 0.52 & 1/19 & 3.89 & 4.77 & 0.82  \\
1/48 & 2.69 & 5.07 & 0.53 & 1/24 & 4.20 & 5.07 & 0.83  \\
1/62 & 2.91 & 5.40 & 0.54 & 1/31 & 4.53 & 5.40 & 0.84  \\
1/78 & 3.11 & 5.69 & 0.55 & 1/39 & 4.84 & 5.69 & 0.85  \\
1/100 & 3.33 & 6.01 & 0.55 & 1/50 & 5.16 & 6.01 & 0.86  \\
1/124 & 3.51 & 6.29 & 0.56 & 1/62 & 5.44 & 6.29 & 0.86  \\
1/158 & 3.72 & 6.60 & 0.56 & 1/79 & 5.75 & 6.60 & 0.87  \\
\hline
\hline
\multicolumn{4}{c|}{DG1} & \multicolumn{4}{c}{DG2} \\
$h$-size & DoA & DoS & DoA/DoS & $h$-size & DoA & DoS & DoA/DoS  \\
\hline
1/16 & 1.75 & 4.52 & 0.39 & 1/11 & 2.75 & 4.56 & 0.60  \\
1/20 & 1.94 & 4.81 & 0.40 & 1/13 & 2.99 & 4.77 & 0.63  \\
1/25 & 2.13 & 5.10 & 0.42 & 1/17 & 3.37 & 5.12 & 0.66  \\
1/32 & 2.34 & 5.42 & 0.43 & 1/21 & 3.68 & 5.40 & 0.68  \\
1/39 & 2.51 & 5.68 & 0.44 & 1/26 & 4.00 & 5.68 & 0.70 \\
1/50 & 2.72 & 6.00 & 0.45 & 1/33 & 4.35 & 5.99 & 0.73 \\
1/63 & 2.92 & 6.30 & 0.46 & 1/42 & 4.71 & 6.30 & 0.75 \\
1/79 & 3.12 & 6.60 & 0.47 & 1/53 & 5.05 & 6.60 & 0.77 \\
\hline
\end{tabular}
\end{table}
}
Looking at the static scaling, we see that since all problems
have an equal number of DoF and the low order methods have a faster computation rate, 
they will finish first, but this time DG1 outperforms CG1 for tetrahedrons. 
We have almost no fall off as the problem size increases and small degradation from every
method as we approach the strong-scaling limit. When we look at the DoE plots
in Figure~\ref{Fig:Test3efficacy}, the higher order methods again dominate the
lower order. The true static-scaling plots confirm that CG2 for tetrahedrons 
is actually the best since it has both the highest DoE and true DoF/s metrics.

\subsection{Test \#4: Different parallel solvers.}
Finally, what happens if we extend the TAS spectrum analysis to a larger-scale computing environment?
Do different parallel solvers/preconditioning strategies affect the performance results? Let us now 
consider PETSc's native finite element library for the CG1 and CG2 methods built on top of 
the DMPlex data structure. Three different multigrid libraries are analyzed for a series of structured 
hexahedron meshes, and the $h$-sizes are once again chosen so that the CG1 and CG2 have 
equal DoF counts. Let us now consider the following analytical solution on a unit cube domain:
\begin{align}\label{Eqn:test4solution}
  u(x, y, z) = 5\sin(6\pi x) \sin(7\pi y) \sin(8\pi z).
\end{align}
Three different parallel multigrid solvers are employed: PETSc's GAMG,
HYPRE's BoomerAMG, and Trilinos' ML. Problem sizes ranging from 
2,048,383 to 133,432,831 DoFs are examined across 32 Haswell nodes (for 
a total of 1024 MPI processes). It should be noted that in the PETSc 
finite element implementation, Dirichlet boundary conditions are 
removed from the system of equations so only the interior nodes are treated
as unknowns for our scaling analyses.
{\tiny
\begin{table}[b]
\center
\caption{Comparison of 3D structured grid (same $h$) CG and DG in PETSc's native finite element
framework\label{Tab:test4}}
\begin{tabular}{cccc|cccc}
\hline
 \multicolumn{4}{c|}{CG1} & \multicolumn{4}{c}{CG2}  \\
$h$-size & DoA & DoS & DoA/DoS & $h$-size & DoA & DoS & DoA/DoS  \\
\hline
1/128 & 1.75 & 6.31 & 0.28 & 1/64 & 3.18 & 6.31 & 0.50 \\
1/160 & 1.94 & 6.60 & 0.29 & 1/80 & 3.47 & 6.60 & 0.53 \\
1/200 & 2.14 & 6.90 & 0.31 & 1/100 & 3.76 & 6.90 & 0.55 \\
1/256 & 2.35 & 7.22 & 0.33 & 1/128 & 4.08 & 7.22 & 0.57 \\
1/320 & 2.54 & 7.51 & 0.34 & 1/160 & 4.38 & 7.51 & 0.58 \\
1/400 & 2.74 & 7.80 & 0.35 & 1/200 & 4.67 & 7.80 & 0.60 \\
1/512 & 2.95 & 8.13 & 0.36 & 1/256 & 4.99 & 8.13 & 0.62 \\
\hline
\end{tabular}
\end{table}
}
\begin{figure}[t]
  \centering
  \subfloat[Mesh-convergence]{\includegraphics[width=0.5\textwidth]{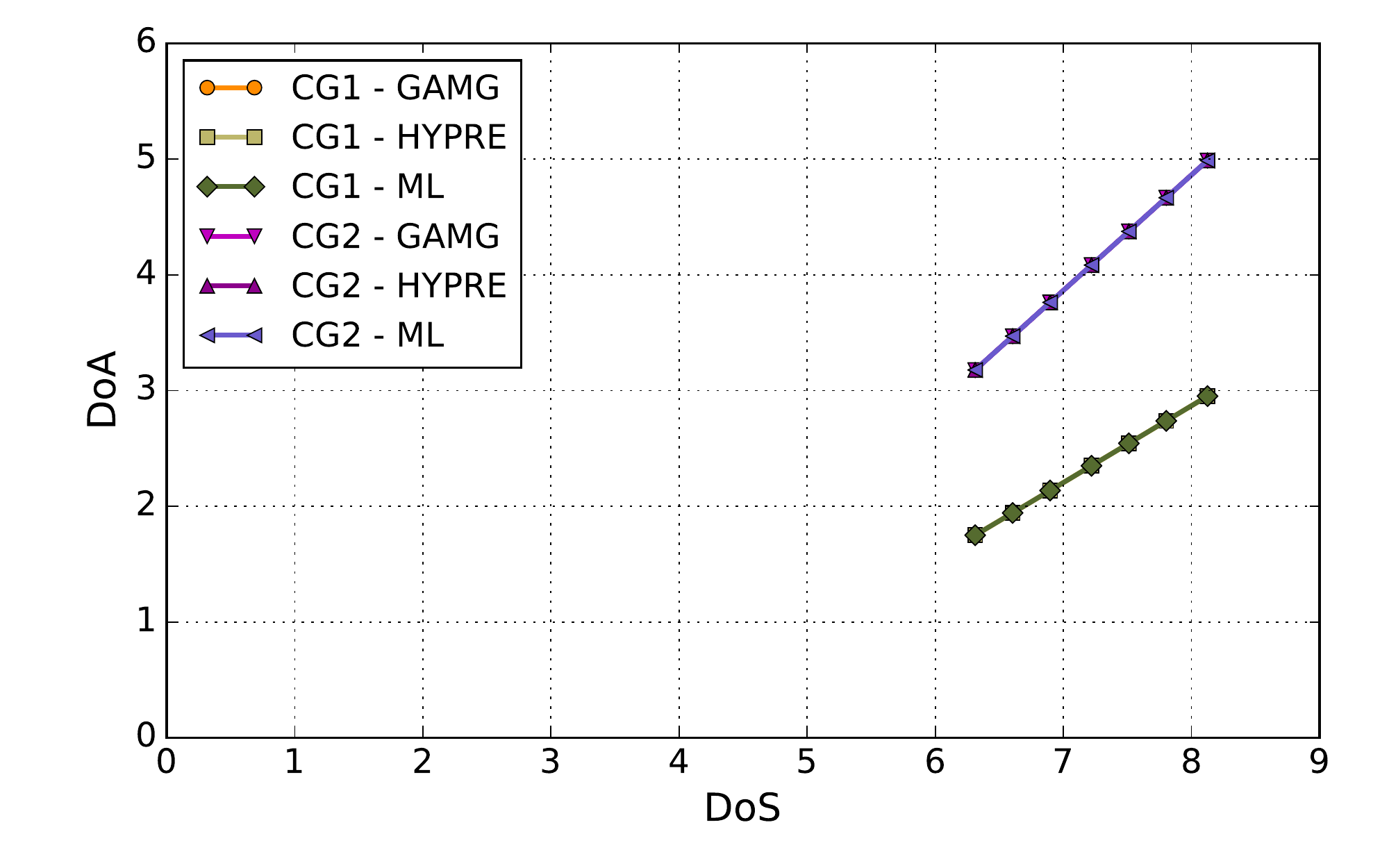}}
  \subfloat[Static-scaling]{\includegraphics[width=0.5\textwidth]{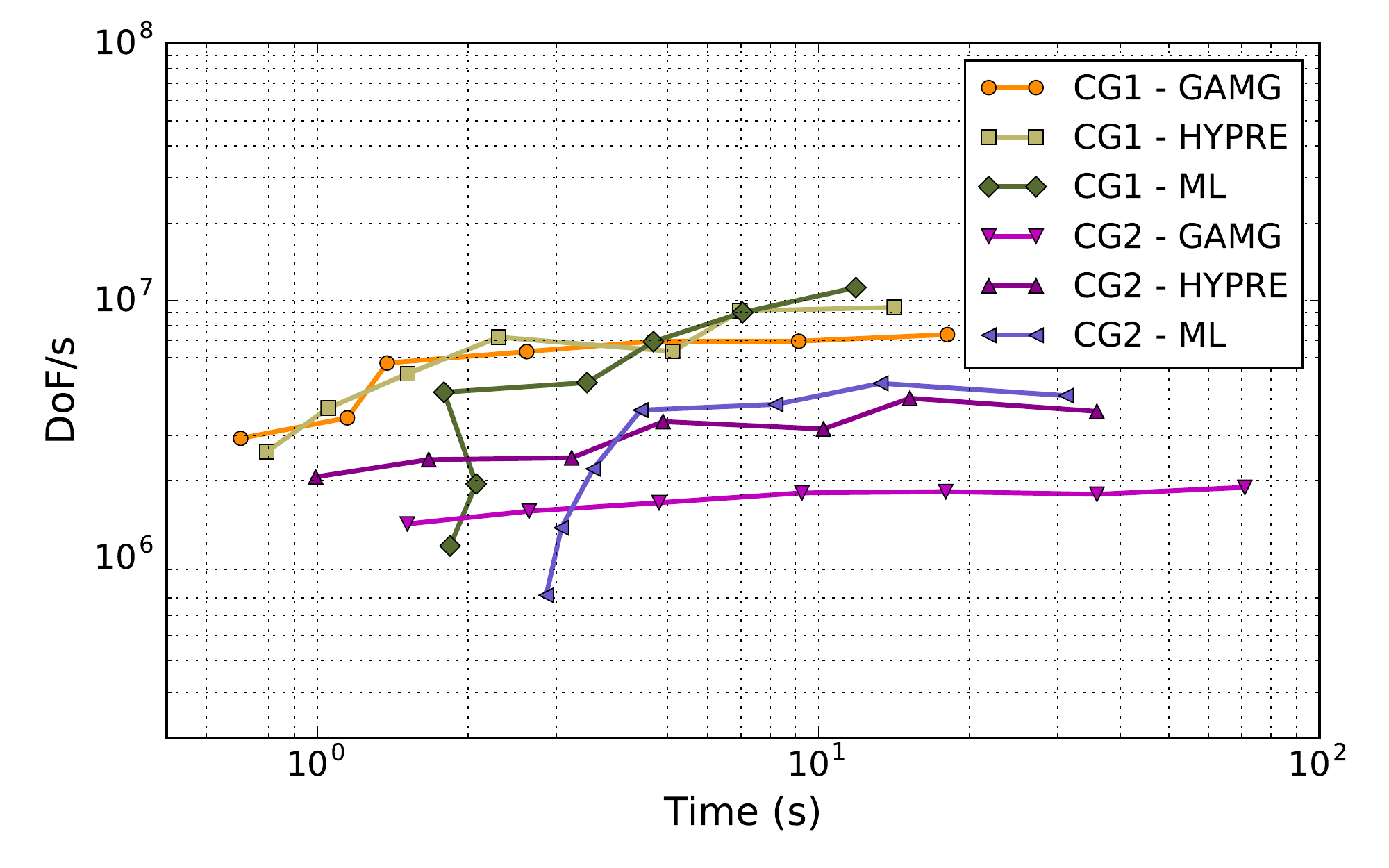}}\\
  \subfloat[DoE]{\includegraphics[width=0.5\textwidth]{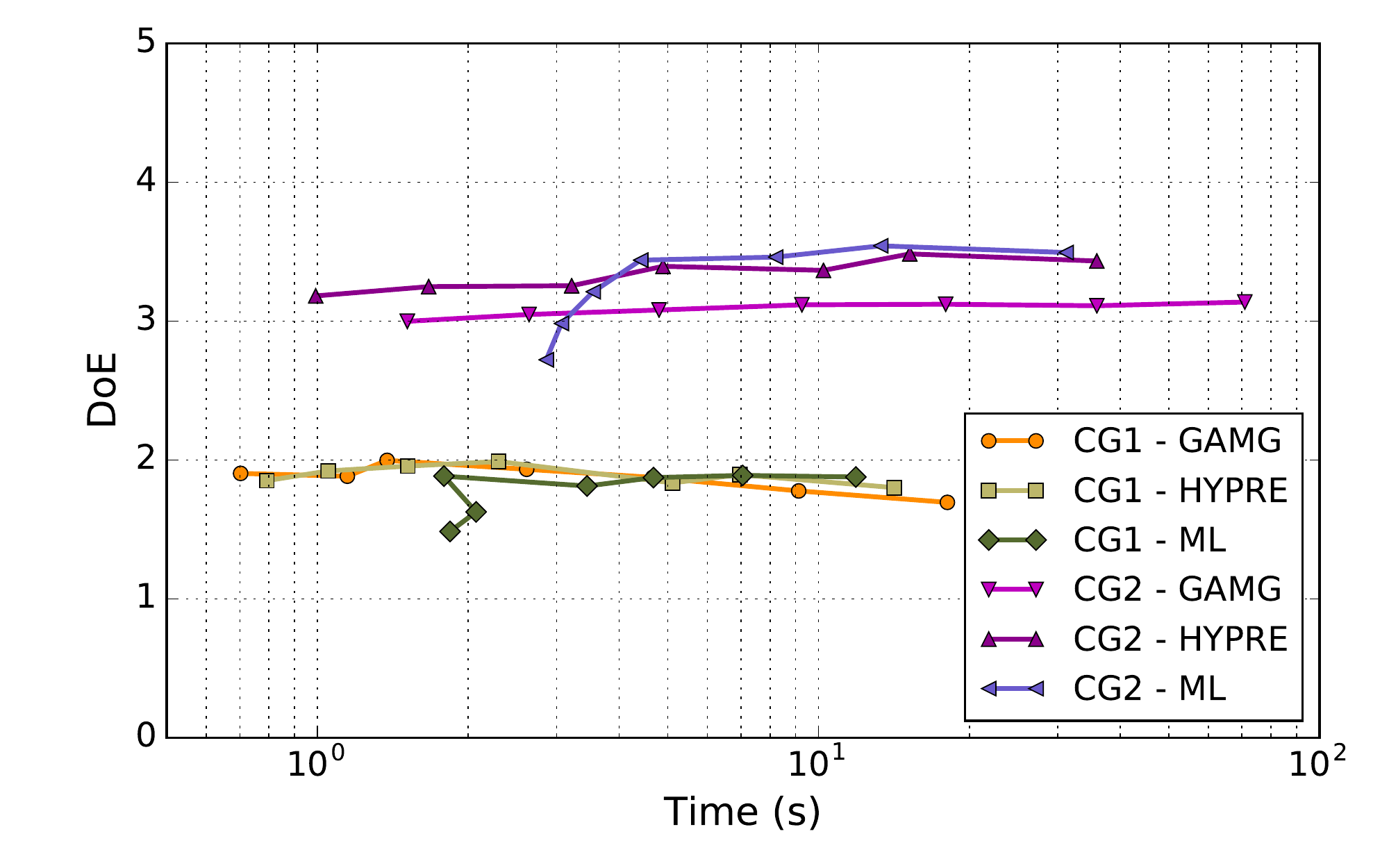}}
  \subfloat[True static-scaling]{\includegraphics[width=0.5\textwidth]{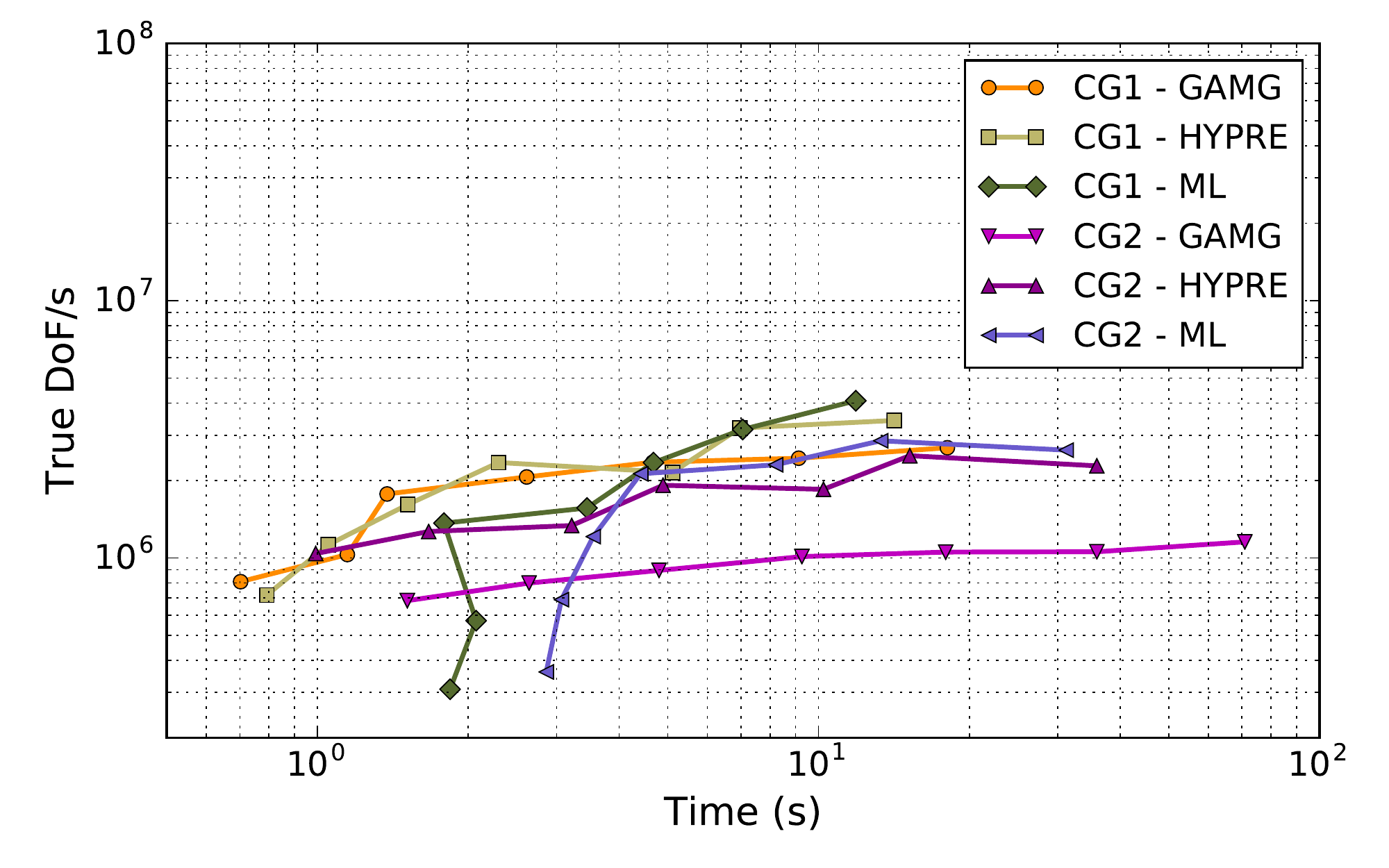}}
  \caption{Test \#4: A complete TAS spectrum analysis of different parallel multigrid solvers across 1024 MPI processes.\label{Fig:Test4}}
\end{figure}

Table \ref{Tab:test4} contains the DoA and DoS information for CG1 
and CG2. As seen from the FEniCS/Dolfin, Firedrake, and deal.II libraries,
the higher order methods have larger DoA/DoS ratios. The complete 
TAS spectrum is shown in Figure~\ref{Fig:Test4}. First, we verify from
the mesh convergence diagram that the solutions obtained from each solver
for a range of $h$-sizes have the expected $\alpha/d$ slopes. Second, the
static-scaling diagram shows how the results for 
each finite element discretization are heavily influenced by the solver. 
Although the ML solver experiences the most significant strong-scaling 
effects, it has the highest peak DoF per second rate for both CG1 and 
CG2. Another observation that can be made from this diagram is that the
GAMG solver has the flattest line, indicating the best scalability in the
strong scaling limit, but it is the least efficient in terms of processing
its DoF. Of course, the data could change if
one were to optimize the solver parameters for this problem. It can be 
also be seen that CG1 has the higher DoF per second rates, but the
DoE diagram indicates that CG2 is actually more efficient. The lines
in this diagram are not completely straight due to the 
strong-scaling effects previously noted. Lastly, it can be seen that both
CG1 and CG2 are grouped closely together in the true 
static-scaling diagrams, suggesting that both discretizations are 
processing the scaled DoFs at an equal pace. Overall, the TAS spectrum analysis 
tells us that CG2 is the more efficient algorithm to use under the 
PETSc DMPlex framework, and the choice of solver would depend on problem size.

\section{Conclusion}\label{sec:conclusion}

By incorporating a measure of accuracy, or convergence of the numerical 
method, into performance analysis metrics, we are able to make meaningful 
performance comparisons between different finite element methods for 
which the worth of an individual FLOP differs due to 
different approximation properties of the algorithm---whether because of 
differences in discretizations, convergence rates, or any 
any other reason. For example, we saw that for the 3D Poisson problem 
with smooth coefficients, the DG1 method may have the highest computation
rate in terms of DoF per time but have low DoE and true DoF per time metrics 
once DoA is taken into consideration. Simultaneously looking at the DoE
and true DoF per time diagrams can further the understanding of how fast
and accurate a particular method is. 
\subsection{Extensions of this work}
The Time-Accuracy-Size (TAS) spectrum analysis opens the door to a variety 
of possible performance analyses. The most logical extension of this work 
would be to to analyze different and more complicated PDEs, but there still exist
some important issues that were not covered in this paper. We now 
briefly highlight some of these important areas of future research:
\begin{itemize}
\item \textit{Arithmetic intensity:} A logical extension of the TAS spectrum 
performance analysis would be to incorporate the Arithmetic Intensity (AI), used in the 
performance spectrum \cite{ChangNakshatralaKnepleyJohnsson2017} and 
roofline performance model \cite{Williams_ACM_2009}. The AI of an algorithm or software is a 
measure that aids in estimating how efficiently the hardware resources and 
capabilities can be utilized. The limiting factor of performance 
for many PDE solvers is the memory bandwidth, so having a high AI
increases the possibility of reusing more data in cache and lowers memory bandwidth 
demands. It can be measured in a number of ways, such as through the 
Intel SDE/VTune libraries or through hardware counters like cache misses.
% RTM: Again, removed registered/trademark symbols -- not appropriate for journal articles, generally.
\item \textit{Numerical discretization:} This paper has solely focused on the 
finite element method using CG and DG discretization, so it would be a worthy
research endeavor to investigate other types of elements like hybrid or mixed elements. 
Furthermore, this type of analysis is easily extendible to other numerical methods like 
the finite difference, finite volume, spectral element, and boundary integral methods.
\item \textit{Accuracy measures:} The accuracy rate metrics were based on the $L_2$ error norm,
but other measures of accuracy or convergence, like the $H^1$ error seminorm, can be used.
The accuracy of different numerical methodologies may sometimes require more than just the standard
$L_2$ error norm. For example, one can validate and verify the performance of finite element methods
for porous media flow models using the mechanics-based solution verification measures described
in \cite{shabouei2016mechanics}.
\item \textit{Solver strategies:} Only the multigrid solvers HYPRE, GAMG, and ML have been
used for the experiments in this paper but there are various other solver and preconditioning strategies 
one can use which may drastically alter the comparisons between the CG and DG
methods. A thorough analysis and survey of all the appropriate solver and preconditioning combinations
may be warranted for any concrete conclusions to be made about these finite element methods. 
There are also different ways of enforcing constraints or conservation laws, solving nonlinear systems with 
hybrid and composed iterations, and handling coefficient jumps in different ways. 
For this, we may also want to incorporate statistics of the
iteration involved~\cite{MorganKnepleySananScott2016}. 
\item \textit{Large-scale simulations:} The computational experiments performed in the previous section
are relatively small, but they can easily scale up so that up to 100K or more MPI processes may be needed.
Furthermore, even if smaller scale comparisons were to be made such as the ones shown in 
this paper, the choice of hardware architecture could play an important role in the scaling analyses.
Intel systems were used to convey some important performance comparisons, but such comparisons
may be very different on systems provided by IBM, AMD, or even NVIDIA.
\end{itemize}

\section*{ACKNOWLEDGMENTS}
The authors acknowledge L.~Ridgway Scott (University of Chicago) and David Ham (Imperial College of
London) for their invaluable thoughts and ideas which improved the overall quality of this paper. 
JC was supported by a grant from the Rice Intel Parallel Computing Center.
MSF would like to acknowledge support from the Ken Kennedy-Cray Inc. Graduate Fellowship Endowment, and Rice Graduate Education for Minorities program.
MGK was partially supported by the U.S. Department of Energy under Contract No. DE-AC02-06CH11357, and by the NSF SI2-SSI 1450339.
R.\ T.\ Mills was supported by the Exascale Computing Project (17-SC-20-SC), a collaborative effort of the U.S. Department of Energy Office of Science and the National Nuclear Security Administration.
This research used resources (the Intel Xeon E5-2698v3 nodes of the Cori Cray XC40 system) of the 
National Energy Research Scientific Computing Center (NERSC), a DOE Office of Science User Facility 
supported by the Office of Science of the U.S. Department of Energy under Contract No. DE-AC02-05CH11231.

\bibliographystyle{siam}
\bibliography{bibliography,petsc,petscapp}
\end{document}